\documentclass[useAMS,usenatbib,usegraphicx]{mn2e}

\usepackage{amssymb}
\usepackage{times}

\title[The dynamical influence on the $m_{\rmn{max}}$--$M_{\rmn{ecl}}$ relation]
{The influence of stellar-dynamical ejections and collisions on the relation between the maximum-star 
and star-cluster-mass}

\author[S. Oh \& P. Kroupa]{Seungkyung Oh\thanks{Member of the International Max Planck Research School
(IMPRS) for Astronomy and Astrophysics at the Universities of Bonn and Cologne.}\thanks{E-mail: 
skoh@astro.uni-bonn.de (SO); pavel@astro.uni-bonn.de (PK)} and Pavel Kroupa\footnotemark[2]
\\
Argelander-Institut f\"ur Astronomie, Auf dem H\"ugel 71, 53121 Bonn, Germany\\
}

\begin{document}

\date{Submitted}

\pagerange{\pageref{firstpage}--\pageref{lastpage}} \pubyear{}

\maketitle

\label{firstpage}

\begin{abstract}
We perform the largest currently available set of direct N-body calculations of young 
star cluster models to study the dynamical influence, especially through the ejections of 
the most massive star in the cluster, on the current relation between 
the maximum-stellar-mass and the star-cluster-mass. 
We vary several initial parameters such as the initial half-mass radius of the cluster, 
the initial binary fraction, and the degree of initial mass segregation. 
Two different pairing methods are used to construct massive binaries for more
realistic initial conditions of massive binaries. 
We find that lower mass clusters ($\leq10^{2.5}~\rmn{M}_{\sun}$) do not shoot out 
their heaviest star. In the case of massive clusters ($\geq1000~\rmn{M}_{\sun}$),
 no most-massive star escapes the cluster within 3~Myr regardless of the initial conditions 
if clusters have initial half-mass radii, $r_{0.5}$, $\geq 0.8$~pc. However, a few of the initially 
smaller sized clusters ($r_{0.5} = 0.3$~pc), which have a higher density, eject their 
most massive star within 3~Myr. 
If clusters form with a compact size and their massive stars are born in a binary system 
with a mass-ratio biased towards unity, the probability that the mass of the most massive star 
in the cluster changes due to the ejection of the initially most massive star can be as large as 20 per cent. 
Stellar collisions increase the maximum-stellar-mass in a large number of clusters
when clusters are relatively dense ($M_{\rmn{ecl}} \geq 10^3~\rmn{M}_{\sun}$ and 
$r_{0.5}=0.3$~pc) and binary-rich. Overall, we conclude that  dynamical effects 
hardly influence the observational maximum-stellar-mass -- cluster-mass relation.
\end{abstract}


\section{Introduction}
Weidner, Kroupa \& Bonnell (2010) compiled from the literature observational data of 100 young star clusters, 
whose masses lie between $\approx 10$ and $\approx 2 \times 10^{5}~\rmn{M}_{\sun}$ and 
whose ages are younger than 4~Myr. They showed that observed young star clusters
exhibit a well-defined correlation between the maximum stellar mass in the cluster,
$m_{\rmn{max}}$, and the mass in stars, $M_{\rmn{ecl}}$, of the cluster. 
An upper age limit of 4~Myr was chosen in order to minimize any evolutionary effects on the sample. 
The examples of evolutionary effects discussed in their paper are as follows. 
Firstly, mass loss of massive stars due to stellar evolution may influence the cluster mass. 
Secondly, gas expulsion leads the cluster to lose a significant amount of its stars 
(i.e. cluster mass) by weakening of the gravitational potential when the residual gas is 
expelled from the cluster. 
However these effects unlikely affect $M_{\rmn{ecl}}$ owing to the young ages of the clusters 
in the \citet{WKB10} sample.
The authors corrected $m_{\rmn{max}}$ for stellar evolution in the case of O-type stars
(note that later than O-type stars would not have evolved much at this young age),
so that the $m_{\rmn{max}}$ values provided in their paper can be considered as initial values.
One process they did not take account of is the dynamical ejection of the most 
massive star from the cluster. The authors commented that it is highly unlikely to happen.
But this has not been studied thoroughly so far. 
Thus it is our aim in this study to investigate how often a young star cluster 
ejects its most massive member.

This observed correlation fits a semi-analytical model well \citep{WK04,WK06,WKB10} which is 
deduced from there being exactly one most massive star in the cluster,
\begin{equation}
1 = \int_{m_{\rmn{max}}}^{m_{\rmn{max}\ast}} \xi(m)\rmn{d}m, 
\end{equation}
subject to the normalisation
\begin{equation}
M_{\rmn{ecl}} = \int_{m_{\rmn{low}}}^{m_{\rmn{max}}} m \xi(m)\rmn{d}m,
\end{equation}
where $m_{\rmn{max}\ast} \approx 150-300~\rmn{M}_{\sun}$ is the fundamental upper limit 
of stellar masses \citep{WK04, WK06, Fi05, OC05, CS10}, $m_{\rmn{low}} = 0.08~\rmn{M}_{\sun}$ 
is the hydrogen burning mass limit (brown dwarfs contribute negligibly to the cluster mass, 
Thies \& Kroupa 2007) and $\xi(m)$ is the stellar initial mass function (IMF).
A pure size-of-sample effect as expected from random sampling has been excluded as 
an origin of the observed correlation.
Details of previous studies on the $m_{\rmn{max}}$--$M_{\rmn{ecl}}$ relation
can be found in \citet{WK04,WK06}, \citet{WKB10} and references therein.
Note that \citet{MC08} argued that at least for the low-$N$ clusters (i.e. low mass clusters)
observed maximum stellar masses do not much deviate from random drawing. This is basically true
but leads the reader to the misinterpretation that a physical origin of the most-massive star
in these clusters is ruled out. But \citet{WKB10} show that in the mass regime considered by \citet{MC08}
a physical $m_{\rmn{max}}$ and a stochastic $m_{\rmn{max}}$ can not be distinguished from each other.
The observed clusters from \citet{WKB10} and their semi-analytical relation are 
reproduced in Fig.~1.  
Interestingly, a similar relation appears in numerical simulations of star 
cluster formation as well. Smoothed particle hydrodynamics numerical simulations 
of massive star formation driven by competitive accretion (Bonnell, Vine \& Bate 2004) showed 
that the most massive star 
in the forming cluster grows following the relation $m_{\rmn{max}}(t) \propto M_{\rmn{ecl}}(t)^{2/3}$ 
with time, t, which is a best-fit to their simulation data. 
This fit agrees with the semi-analytical $m_{\rmn{max}}$--$M_{\rmn{ecl}}$ relation very well for 
clusters with $M_{\rmn{ecl}}\lesssim10^3~\rmn{M}_{\sun}$. 
Furthermore, fragmentation-induced starvation studied with radiation-hydrodynamical simulations 
of massive star formation using the adaptive-mesh code 
FLASH \citep{Pet10} also reproduce the relation found by \citet{BVB04}. 
Data from the numerical simulations of star cluster formation including the two studies mentioned above are 
also plotted in Fig.~1. The simulation data are in good agreement with the observed data.

Although most of the clusters follow the relation well, there is a spread of 
$m_{\rmn{max}}$ values at a given cluster mass. Is this spread due to stochastic 
effects that occur during the formation of a cluster, or does it mask a true 
physical functional dependence of $m_{\rmn{max}}$ on $M_{\rmn{ecl}}$?
Dynamical processes can exert an influence on the relation
during the early evolution of the cluster.
Stars can be dynamically ejected through energetic few body interactions. 
The lightest star among the interacting stars generally obtains the highest 
velocity and it is unlikely that the most massive star is ejected.  
Several theoretical studies, nevertheless, have shown that massive stars can be 
dynamically ejected under certain circumstances such as from a small group of 
massive stars lacking low-mass stars \citep{CP92,GGP09,GG11,FP11}, through binary-single 
\citep{HF80} and binary--binary interactions \citep{LD90}. 
The high efficiency of dynamical ejections from dense stellar systems can indeed
explain the difference between the observed and expected number of OB-type stars 
in the Orion Nebula Cluster \citep{PK06}. 
Furthermore, dense and massive R136-type clusters are efficient in expelling massive stars 
(Banerjee, Kroupa \& Oh 2012).
Thus it may be possible that the heaviest star in a cluster be dynamically ejected from the cluster.

In this contribution we assume there exists an exact function, 
$m_{\rmn{max}} = \rmn{fn}(M_{\rmn{ecl}})$,
and we study the ejection of the heaviest star in a cluster using 
direct N-body integration to investigate the effect on 
the $m_{\rmn{max}}$--$M_{\rmn{ecl}}$ relation. 
Details of the initial conditions of the cluster models and of the calculations 
are described in Section 2 and then results are shown in Section 3. 
The discussion and the conclusions follow in Sections~4 and ~5.

\begin{figure}
 \centering
 \includegraphics[width=85mm]{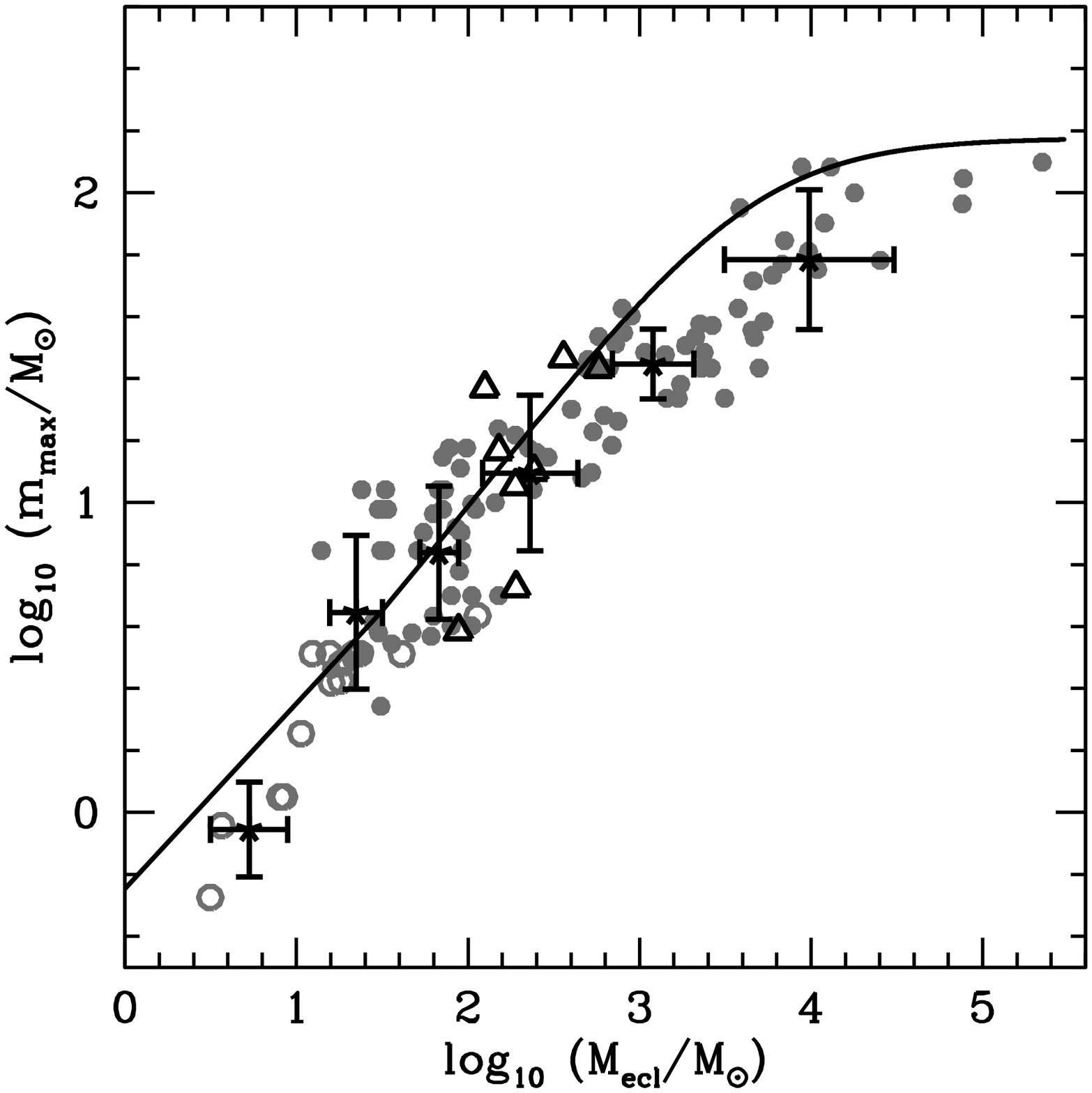}
 \caption{Mass of the most massive star versus cluster mass from observational data in \citet{WKB10}. 
  Each grey filled circle is a star cluster from their table~B1. Black solid line is the semi-analytical 
  $m_{\rmn{max}}$--$M_{\rmn{ecl}}$ relation from equations~1 and~2  
  assuming an upper limit of the stellar mass of $150~\rmn{M}_{\sun}$ \citep{WK04,WK06}.
  The open circles are the mass of the most massive member in the group vs 
  the total group mass in 14 young stellar groups in Taurus, Lupus3, ChaI, and IC348 
  \citep{KM11}. These young (low-mass) stellar groups also follow the 
  $m_{\rmn{max}}$--$M_{\rmn{ecl}}$ relation well. 
  Black points and errorbars represent the average and standard deviation of 
  $\log_{10} M_{\rmn{ecl}}$ and $\log_{10} m_{\rmn{max}}$ in each bin each of which contains 22 clusters 
  for the upper 5 bins and 4 clusters for the lowest cluster mass bin.
  Data from several numerical simulations of star formation 
  (Bonnell, Vine \& Bate 2004; Bate 2009, 2012; Smith, Longmore \& Bonnell 2009; Peters et al. 2010)
  are included in the figure as open triangles.}
\end{figure}

\section{Models}
We perform a large set of direct N-body calculations of young star clusters
using {\sevensize NBODY}6 \citep{Aa99} with various initial conditions.
Cluster masses range from $10~\rmn{M}_{\sun}$ to $10^{3.5}~\rmn{M}_{\sun}$ with an interval
of 0.5 on the logarithmic scale and each mass is initialized with two different half-mass 
radii, $r_{0.5} = 0.3$~pc and $0.8$~pc.
To study the effect of binaries, we adopt two extreme binary fractions which are 
0 (all stars are single) and~1 (all stars are in binary systems). 
The initial binary population used in this study is described in Section 2.1.
Single star clusters (all stars are single) are chosen for comparison purpose 
only since most stars in actuality form in a binary system \citep{GK05}. 

For each cluster mass the number of stars, $N_{\rmn{star}}$, is assigned by dividing 
the cluster mass by the average stellar mass of the cluster
\begin{equation}
 N_{\rmn{star}} = \frac{M_{\rmn{ecl}}}{<m>},
\end{equation}
where the average stellar mass of the cluster, $<m>$, is
\[
<m> = \frac{\int_{m_{\rmn{low}}}^{m_{\rmn{max}}}m\xi(m)\rmn{d}m}
       {\int_{m_{\rmn{low}}}^{m_{\rmn{max}}}\xi(m)\rmn{d}m}.
\]
We adopt the canonical two-part power law IMF \citep{Kr01,Ket12}, 
\begin{equation}
 \xi(m) \propto m^{-\alpha_{i}},
\end{equation}
where
\[
 \begin{array}{l}
 \alpha_{1} = 1.3, \quad 0.08 \le m/\rmn{M}_{\sun} < 0.50, \\
 \alpha_{2} = 2.3, \quad 0.50 \le m/\rmn{M}_{\sun}\le m_{\rmn{max}}.
 \end{array}
\]
We use $m_{\rmn{max,WK}}\equiv m_{\rmn{max}}$, which is calculated using 
the semi-analytical relation (equations~1 and~2) assuming the fundamental 
upper limit of stellar masses to be $m_{\rmn{max}\ast} = 150~\rmn{M}_{\sun}$. 
The solid line in Fig.~1 represents $m_{\rmn{max,WK}}$.

Individual masses of all stars but one (i.e. $N_{\rmn{star}}-1$ stars) 
in each cluster are randomly drawn from the IMF (equation~4) with a stellar mass range 
from $0.08~\rmn{M}_{\sun}$ to $m_{\rmn{max,WK}}$. To simplify the analysis, 
one $m_{\rmn{max,WK}}$ star is added so that every cluster has at least 
one star with a mass of $m_{\rmn{max,WK}}$. 
\footnote{This procedure is not required if optimal sampling of stellar masses from the IMF 
\citep{Ket12} were used instead of random sampling. Optimal sampling was not available 
though at the time the present library of clusters was computed.} 
This procedure removes the stochastic effects on the initial $m_{\rmn{max}}$. 
This choice, however, gives a bump at the most massive mass-bin 
in the IMF of the cluster, which is especially significant in small-$N$ (i.e. low-mass) clusters.
As the dynamical ejection of massive stars occurs by close encounters between massive stars 
\citep{CP92,LD90}, our enforcement of having a $m_{\rmn{max,WK}}$ star may enhance the ejections
at a given $M_{\rmn{ecl}}$ by overpopulating massive stars. 
We find that this choice would not change our conclusion (see further discussion in Section~4).

Positions and velocities of stars in the cluster are generated according to the
Plummer model \citep{AHW74} which is the simplest stationary solution of 
the Collisionless Boltzmann Equation \citep{HH03,Kr08} and is an excellent description 
of the nearest star cluster, the Hyades \citep{Ret11}.  
To study the effect of initial mass segregation, half of our cluster models
are initially mass segregated. The method for constructing initially mass segregated 
clusters in which positions and velocities are dependent on stellar masses is described 
in Section 2.2. For unsegregated clusters positions and velocities are assigned to stars 
independently of their masses. 

Dynamical time scales such as the crossing time and the median two-body relaxation time
are important tools to estimate the dynamical evolution of stellar systems.
The initial crossing time is
\begin{equation}
 t_{\rmn{cr}} = {{2 r_{0.5}} \over {\sigma}},
\end{equation}
where $\sigma$ is initial velocity dispersion, $\sigma = \sqrt{G M_{\rmn{ecl}}/r_{\rmn{grav}}}$,
$r_{\rmn{grav}}\approx2.6~r_{0.5}$ is the gravitational radius \citep{BT87,Kr08}. The relaxation time is
\begin{equation}
 t_{\rmn{rel}} = 0.1{{N_{\rmn{star}}} \over {\ln N_{\rmn{star}}}} t_{\rmn{cr}} .
\end{equation}

Initial conditions of all cluster models are listed in Table 1.
Table 2 shows the physical properties of the 6 different-mass clusters.
All clusters are evolved up to 5~Myr. And stellar evolution is taken into account 
using the stellar evolution library (Hurley, Pols \& Tout 2000) in the {\sevensize NBODY}6 code. 
We carry out 100 computations for each set of initial conditions. 
In total 7200 models are thus calculated with a standard PC. 
In addition, we perform 10 calculations for clusters with $M_{\rmn{ecl}}=10^4~\rmn{M}_{\sun}$ 
and with the same initial conditions as MS3OP in Table~1. We compute only the most energetic cluster model (MS3OP) 
in our library for $M_{\rmn{ecl}}=10^{4}~\rmn{M}_{\sun}$ because of the high computational cost for massive clusters.
This is the largest currently existing systematically generated library of young star cluster models. 

\begin{table}
 \centering
 \caption{Initial conditions of cluster models. Model name is in the first column. 
  Column 2 indicates the initial mass segregation, N standing for an initially 
  unsegregated cluster while Y signifies an initially segregated cluster. 
  Columns 3 and 4 present the initial half-mass radius of the cluster, $r_{0.5}$, 
  and the initial binary fraction, $f_{\rmn{bin,i}}$, respectively.
  The OP and RP in the $f_{\rmn{bin,i}}$ column represent the pairing method 
  for the massive binaries: ordered pairing, and random pairing, respectively. 
  The description of the pairing methods can be found in Section 2.1.}
 \begin{tabular}{@{}lccc@{}}
  \hline
  Model & mass segregation & $r_{0.5}$ [pc] & $f_{\rmn{bin, i}}$\\
  \hline
  NMS3S & N & 0.3 & 0    \\
  NMS3RP & N & 0.3 & 1 (RP) \\
  NMS3OP & N & 0.3 & 1 (OP) \\
  NMS8S & N & 0.8 & 0    \\
  NMS8RP & N & 0.8 & 1 (RP) \\
  NMS8OP & N & 0.8 & 1 (OP) \\
  MS3S  & Y & 0.3 & 0   \\
  MS3RP  & Y & 0.3 & 1 (RP) \\
  MS3OP  & Y & 0.3 & 1 (OP) \\
  MS8S  & Y & 0.8 & 0   \\
  MS8RP  & Y & 0.8 & 1 (RP) \\
  MS8OP  & Y & 0.8 & 1 (OP) \\
 \hline
 \end{tabular}
\end{table}

\begin{table*}
 \caption{Characteristics of clusters with different sizes and masses. 
 Cluster mass, $M_{\rmn{ecl}}$, number of stars in a cluster, $N_{\rmn{star}}$, 
 initial mass of the most massive star in the cluster, $m_{\rmn{max,i}}$ (from equations~1 and~2), 
 and the tidal radius, $r_{\rmn{tid}}$, are presented in columns 1-4. The $r_{\rmn{tid}}$ is 
 obtained from $r_{\rmn{tid}} = R_{\rmn{GC}}(M_{\rmn{ecl}}/(3M_{\rmn{gal}}))^{1/3}$ by assuming 
 $M_{\rmn{gal}}$, the Galactic enclosed mass within the galactocentric distance ($R_{\rmn{GC}}$)
 of 8.5~kpc, to be $5 \times 10^{10}~\rmn{M}_{\sun}$. 
 The initial crossing, $t_{\rmn{cr}}$, and the initial relaxation time, $t_{\rmn{rel}}$, for two 
 different cluster sizes, $r_{0.5} =$ 0.3 and 0.8~pc, are given in columns 5-8. 
 Each cluster model in Table 1 contains all six different mass clusters in this table. 
 In total thus 72 different initial cluster configurations are used in this study, 
 whereby 100 random realisations of each configuration are computed with Aarseth's {\sevensize NBODY}6. 
 The last model ($M_{\rmn{ecl}}=10^4~\rmn{M}_{\sun}$) is only computed 10 times and for the MS3OP configuration. }
 \begin{tabular}{@{}lccccccc@{}}
 \hline
 $M_{\rmn{ecl}}$ [M$_{\sun}$] & $N_{\rmn{star}}$ & $m_{\rmn{max,i}}$ [M$_{\sun}$] &
 $r_{\rmn{tid}}$ [pc]  
 &\multicolumn{2}{c}{$t_{\rmn{cr}}$[Myr]}
 &\multicolumn{2}{c}{$t_{\rmn{rel}}$[Myr]} \\
 & & & &($r_{0.5}= 0.3$pc & $0.8$pc & $0.3$pc & $0.8$pc)\\
 \hline
 $10^{1.0}$ & 28   & 2.1  & 3.5 & 2.56 & 11.13 & 2.14 &  9.35 \\
 $10^{1.5}$ & 76   & 4.5  & 5.1 & 1.44 &  6.26 & 2.52 & 10.98 \\
 $10^{2.0}$ & 214  & 9.7  & 7.4 & 0.81 &  3.52 & 3.22 & 14.03 \\
 $10^{2.5}$ & 618  & 21.2 & 10.9& 0.45 &  1.98 & 4.37 & 19.03 \\
 $10^{3.0}$ & 1836 & 43.9 & 16.0& 0.26 &  1.11 & 6.24 & 27.19 \\
 $10^{3.5}$ & 5584 & 79.2 & 23.5& 0.14 &  0.63 & 9.30 & 40.50 \\
 $10^{4.0}$ & 17298& 114.7& 34.5& 0.08 &       & 14.32&       \\ 
 \hline
 \end{tabular}
\end{table*}

\subsection{Primordial binaries}
To set up the primordial binaries we require their initial orbital parameters 
such as periods, eccentricities, and mass-ratios. 
The initial period distribution adopted in this study is equation (8) in \citet{Kr95b}, 
\begin{equation}
f_{P} = 2.5{{{\log_{10} P - 1}} \over {45 + (\log_{10} P - 1)^{2}}},
\end{equation}
where period, $P$, is in days. With this distribution function, 
minimum and maximum log-periods, $\log_{10}~P_{\rmn{min}}$ 
and $\log_{10}~P_{\rmn{max}}$, are 1 and 8.43, respectively. 
This function shows a flat distribution at long-periods and is in good
agreement with the period distribution of low-density young stellar aggregates such as 
Taurus-Auriga \citep{KP11, MKO11}.
The initial eccentricity distribution follows the thermal distribution, $f(e)=2e$ \citep{Kr08}.
Pre-main-sequence eigenevolution \citep{Kr95b} is not included in our calculations
as our emphasis is on the massive stars.

Initial mass-ratios of low-mass stars such as G-, K- and M-dwarf binaries can be well described 
with random pairing \citep{Kr95a,Kr08}. Observational studies of OB-type binaries, 
however, show that they tend to have similar mass companions \citep{GM01,Set08,SGE09}. 
\citet{SE11} show that the mass-ratio distribution of O star binaries seems uniform 
in the range $0.2 \le m_{2}/m_{1} \le 1.0$, $m_{1}$ being the primary and $m_{2}$ the secondary 
mass. In any case the mass-ratios of massive binaries are high compared to low-mass stars.
Random pairing cannot produce the observed mass-ratio distribution of massive binaries 
since it typically leads to massive stars being paired with low mass stars which are the majority 
in the cluster. Thus random pairing over the whole stellar mass-range for OB star primaries is 
ruled out by the observation. A different pairing method for masses of binary components 
is needed to create massive binaries.
 
In this study we introduce a simple method to generate massive binaries
having mass-ratios biased towards unity.
First, all stellar masses are randomly drawn from the IMF, and then, stars
more massive than $5~\rmn{M}_{\sun}$ are sorted with decreasing mass and 
the others are retained in random order. 
We pair stellar masses in order so that a massive star has the next massive one 
as a companion, while stars less massive than 5 M$_{\sun}$ have a companion
which is randomly distributed.
Thus binaries with primary masses more massive than 5 M$_{\sun}$
have mass-ratios biased towards unity. We call this method "ordered pairing" (OP).
Note that star clusters with masses $\lesssim$ 100 M$_{\sun}$ contain no stars 
more massive than $5~\rmn{M}_{\sun}$ in this paper, 
thus OP clusters with $M_{\rmn{ecl}}\leq10^{1.5}~\rmn{M}_{\sun}$ are the same 
as random pairing (RP) clusters.
For a deep discussion on pairing methods for binaries, 
see \citet{Ket09} and Weidner, Kroupa \& Maschberger (2009) who study several pairing mechanisms. 

\subsection{Primordial mass segregation}
Many young star clusters exhibit evidence for mass segregation \citep{Get04,CGZ07}. 
It has been under debate whether the observed mass segregation of young star clusters 
is the outcome of the star formation processes or of dynamical evolution of the clusters
since a certain time is needed for it to occur dynamically. Some observed clusters seem
too young for mass-segregation to have occurred. 
The dynamical mass segregation time scale is
\begin{equation}
t_{\rmn{ms}} \approx {{<m>} \over {m_{\rmn{massive}}}} t_{\rmn{rel}},
\end{equation}
where $m_{\rmn{massive}}$ is the mass of the massive star. 
For some clusters, $t_{\rmn{ms}}$ could be
shorter than or comparable to their age (e.g. the Orion Nebula Cluster which has 
$t_{\rmn{ms}}$ of about 0.1~Myr, Kroupa 2002).
Thus it is difficult to determine whether an observed mass segregation is primordial or 
the result of dynamical evolution. To study the influence of primordial mass segregation 
on the early dynamical evolution of clusters would give a hint for an answer to this problem.
However it is beyond the scope of this paper as a deeper study on individual clusters
is required to do that.  

It is expected that initially mass segregated clusters ought to be more efficient 
in ejecting massive stars thus allowing the distribution of massive stars to be used 
as a constraint on the issue of initial mass segregation \citep{CP92,GB08,Get11}.
In order to create mass-segregated clusters we use the method introduced in 
Baumgardt, de Marchi \& Kroupa (2008). 
Details of setting up the segregated cluster are described in their Appendix. 
With this method, the heaviest star is most bound to the cluster and is located in 
the core of the cluster.  
And the cluster is initially in virial equilibrium and follows the Plummer density profile.  
Although one can vary the degree of mass segregation with this method,
for simplicity, mass-segregated model clusters in this study are fully segregated.
Thus the segregated clusters have a core of massive stars
in the centre of the cluster at the beginning of cluster evolution. 

We stress that such N-body models of initially fully mass-segregated clusters with a 100
per cent binary fraction and a mass-ratio near unity for the massive binaries have 
never been attempted before. 

\subsection{The $N$-body code}
{\sevensize NBODY}6 is a fully collisional $N$-body code which calculates the force 
on a particle from other particles with direct summation. 
It uses the Hermite scheme for integrating the orbits of stars. 
The code adapts individual time steps depending on the local environment and 
the Ahmad-Cohen neighbour scheme \citep{AC73} for calculation efficiency. 
For treating close encounters, Kustaanheimo-Stiefel two-body regularization and Chain regularization 
for higher order multiple systems are used. 

Stellar evolution is implemented in the code using fitting functions
with the Single Star Evolution package \citep{HPT00} 
and the Binary Star Evolution package (Hurley, Tout \& Pols 2002), which allows a collision
between two components of a binary. Details of stellar evolution in {\sevensize NBODY}6 can 
be found in \citet{Hu08}.
Since we activate stellar evolution in the code, a star has a radius instead of being 
a point mass particle and may collide with other single stars or its companion in a binary system 
by close encounters and/or binary hardening.
When they merge, a mass of one star is replaced by the sum of the colliding stars
and the other star is replaced by a massless particle with a large distance so that
it is removed from the calculation as a massless escaper.
A metallicity of $Z = 0.02$ (solar) is adopted in all our calculations.

\section{Results}
Clusters keep losing their mass due to stars escaping from them via the two-body 
relaxation process or through dynamical ejections besides stellar evolution, and 
so we only count stars found inside the tidal radius as cluster members.
A cluster mass does not change much (at most $\approx3$ per cent on average by 3~Myr) 
within the first few Myr. Therefore the change of the $m_{\rmn{max}}$--$M_{\rmn{ecl}}$ 
relation within the first few Myr is mainly caused by the change of $m_{\rmn{max}}$.
There are three ways to change the maximum stellar mass as a cluster 
evolves. First, stellar evolution changes the mass of the heaviest star in the cluster. 
Stars lose their mass with time via stellar winds. The mass loss rate is 
dependent on the stellar mass. The more massive a star is the larger its mass-loss rate 
is. In the case of the most massive cluster in our model ($M_{\rmn{ecl}}=10^{3.5} 
~\rmn{M}_{\sun}$), its heaviest star with an initial mass of $\approx 80~\rmn{M}_{\sun}$
loses mass to become a $\approx64~\rmn{M}_{\sun}$ star at 3~Myr.
On the other hand for clusters with $M_{\rmn{ecl}}\leq10^{3}~\rmn{M}_{\sun}$, 
$m_{\rmn{max}}$ remains almost the same as the initial value since the heaviest 
stars in such clusters do not evolve much in a few Myr.
Secondly, stars can physically collide. If two stars collide and become a more massive 
star than the initially heaviest star, or if the initially heaviest star collides with 
another star, then the new heaviest star will lie off the initial relation. 
Massive stars generally move into the cluster centre which has a high stellar 
density. Thus a massive star may collide with another massive star in the cluster centre. 
Lastly, dynamical ejection of the initially heaviest star 
in the cluster also changes $m_{\rmn{max}}$ by replacing it with the initially second 
massive star in the cluster. 

Stellar evolution and dynamical ejection leads to the cluster having a smaller maximum stellar mass,
while stellar collisions increase the maximum stellar mass in the cluster.
In our models all three of these effects occur.  However here we concentrate only on stellar collisions 
and ejections as the initial masses of the heaviest stars are set to be equal 
for the same cluster mass. Thus stellar evolution does not produce the $m_{\rmn{max}}$ spread at 
the same cluster mass. And, we are particularly interested how often the heaviest star is ejected 
from the clusters.

Using direct N-body calculations, we study the dynamical effect on the 
$m_{\rmn{max}}$--$M_{\rmn{ecl}}$ relation with various initial conditions of 
the clusters (Tables~1 \&~2). 
Although the clusters are evolved up to 5~Myr, we only use the results up to 3~Myr 
since stellar evolution begins to play a dominant role in changing the value 
of $m_{\rmn{max}}$ of the $10^{3.5}~\rmn{M_{\sun}}$ cluster at around 3.5~Myr.  
The initially most massive star of the $10^{3.5}~\rmn{M_{\sun}}$ cluster 
becomes a blackhole after 4~Myr. Furthermore, most of the observed clusters used in the study of 
the $m_{\rmn{max}}$--$M_{\rmn{ecl}}$ relation \citep{WKB10} are younger than 3~Myr.

For clarification, we refer to the heaviest star in the cluster at 0~Myr as $\rmn{S_{MAXI}}$ 
and to the mass of the most massive star in the cluster at a given snapshot as $m_{\rmn{max}}$.

\begin{figure*}
 \centering
 \includegraphics[width=140mm]{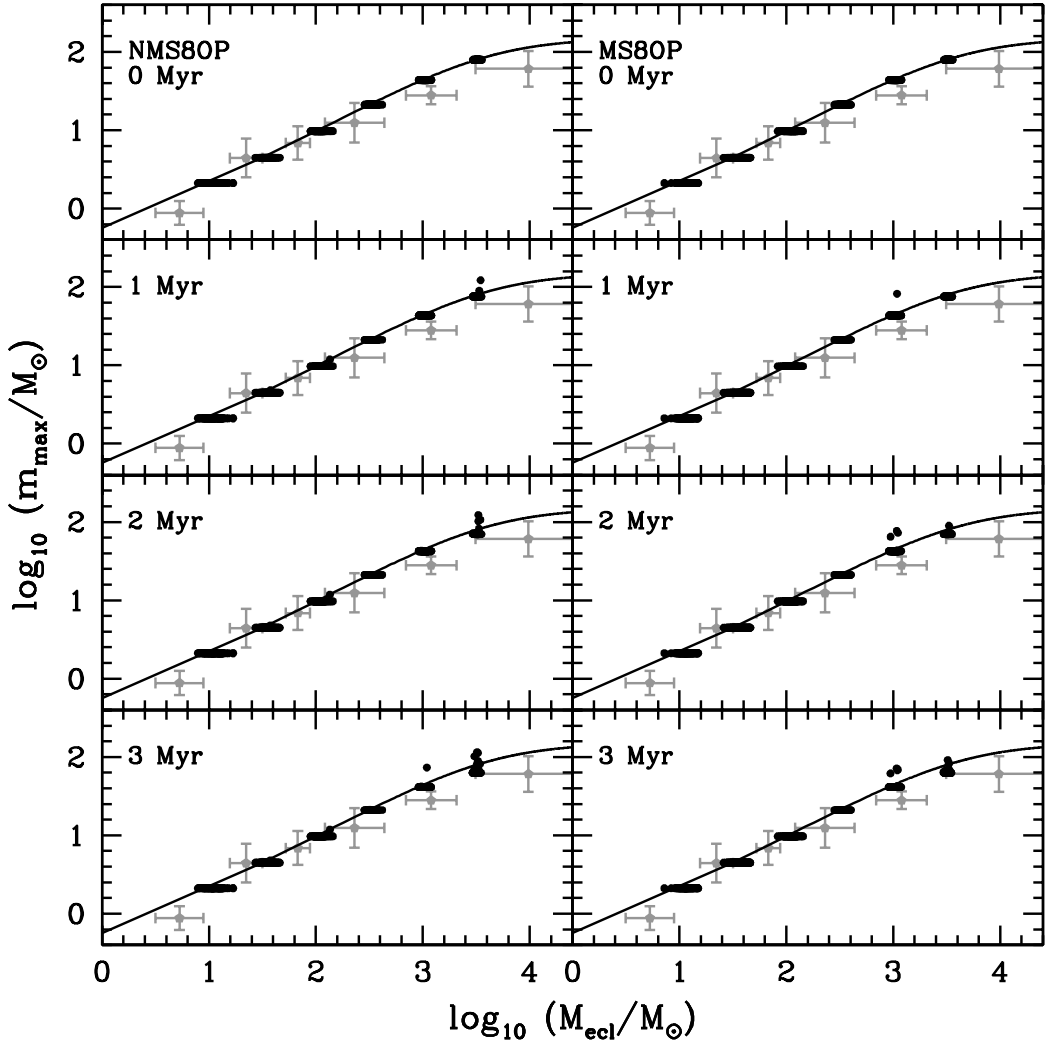}
 \caption{
 The mass of the most-massive star versus the cluster mass for the unsegregated ({\it left}: NMS8OP) and
 the segregated ({\it right}: MS8OP) clusters with $r_{0.5}=0.8$~pc and massive binaries with mass-ratios
 biased towards unity at 0, 1, 2, and 3~Myr (from top to bottom). Each black point indicates a cluster.
 The grey data are the average observational data as in Fig.~1.
 The solid line represents the semi-analytical model from \citet{WK04,WK06}
 assuming an upper limit of stellar mass of $150~\rmn{M}_{\sun}$ (equations~1 and~2).
 As we fix the number of stars at a certain cluster mass, the cluster mass slightly varies for
 each realization for the same cluster mass model.
 We only plot the result up to 3~Myr since the initially heaviest star of the cluster with
 $10^{3.5}~\rmn{M}_{\sun}$ loses a large amount of its initial mass within 3.5~Myr, thereafter
 stellar evolution affects the relation. The stars that appear above the solid curve are merger products.
 }
\end{figure*}

\begin{figure*}
 \centering
 \includegraphics[width=140mm]{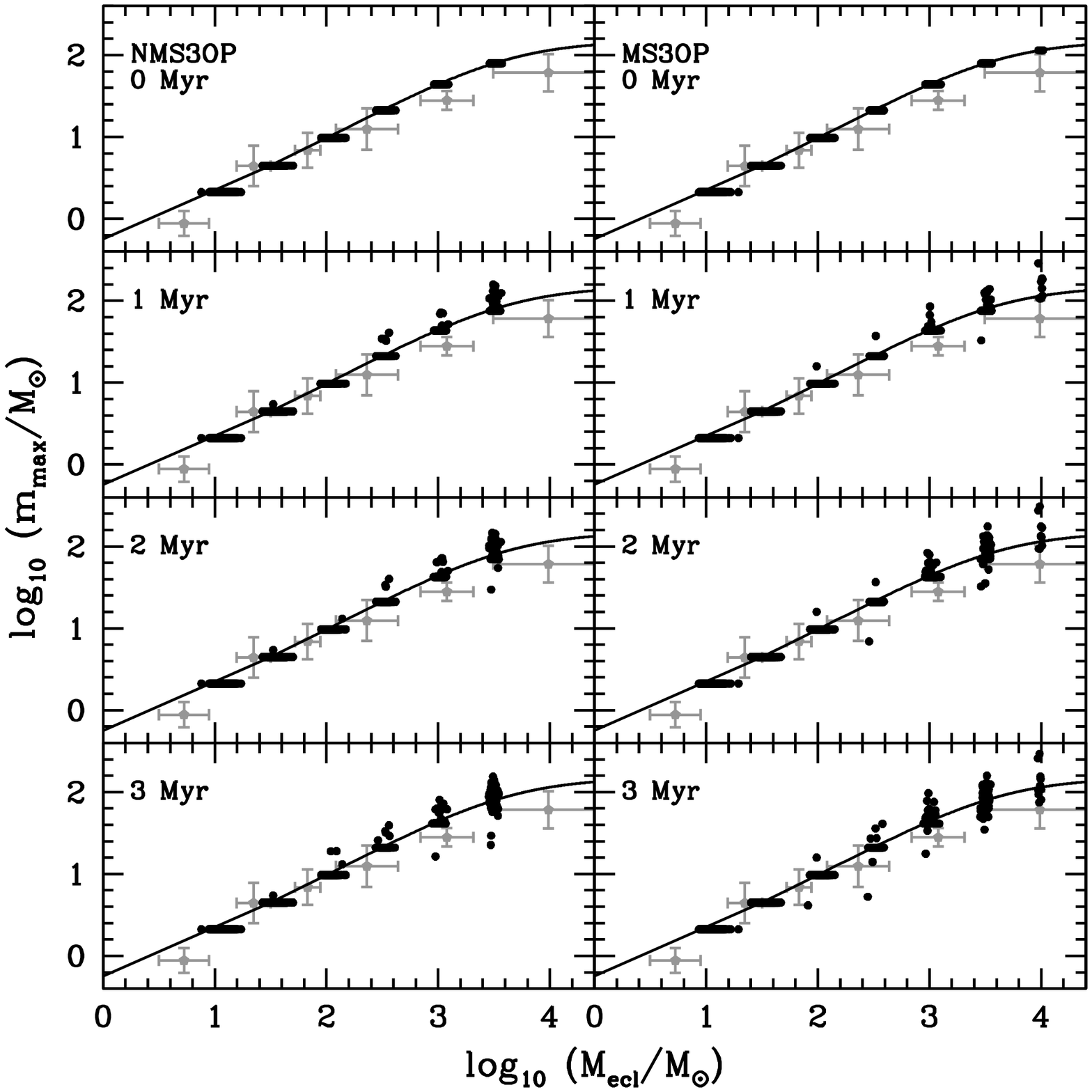}
 \caption{Same as Figs 1 and 2, but for the OP clusters with $r_{0.5} = 0.3$~pc
 ({\it left}: NMS3OP, {\it right}: MS3OP).
 Note how including OP binaries and mass segregation increases the scatter in the diagrams.
 }
\end{figure*}

\subsection{Dynamical ejection of $\rmn{S_{MAXI}}$}
Stars can be ejected from a cluster via close encounters between a hard binary and
a single star/binary. During the encounter, the hard binary gives its binding energy
to the interacting star/stars in the form of kinetic energy and it hardens.
The star that gained kinetic energy may be ejected from the cluster.
Generally the lightest one among the interacting stars attains the highest velocity
after the interaction. Thus the dynamical ejection of massive stars preferably occurs 
from interactions between massive stars. \citet{LD90} showed that binary-binary 
interactions are the most efficient way for producing massive runaways.

We consider $\rmn{S_{MAXI}}$ to be dynamically ejected if the star is further away 
from the cluster centre than the tidal radius of the cluster. 
The number of clusters which eject their $\rmn{S_{MAXI}}$, $N_{\rmn{resc}}$, is 
listed in Table~3 for all cluster models. 

In the following subsections, results on the dynamical ejection of $\rmn{S_{MAXI}}$ 
are discussed separately for models with different initial half-mass radii.

\subsubsection{The $r_{0.5} = 0.8$~pc models}
In Table~3 there are two clusters with $r_{0.5}=0.8$~pc whose $\rmn{S_{MAXI}}$  
is located further than the tidal radius of the cluster at 3~Myr. 
The most massive star of one not initially mass-segregated $10~\rmn{M}_{\sun}$ cluster was, 
in fact, located outside of the tidal radius of the cluster, which is $\approx 3.5$~pc, 
at 0~Myr. Therefore this case is not due to dynamical ejection. 
Thus only 1 out of 3600 model clusters with $r_{0.5} = 0.8$~pc eject their $\rmn{S_{MAXI}}$ 
within 3~Myr.

Fig.~2 shows the $m_{\rmn{max}}$--$M_{\rmn{ecl}}$ relation of the binary-rich clusters 
with $r_{0.5}=0.8$~pc and massive binaries paired by the OP method (models~NMS8OP and~MS8OP) 
at different ages of the clusters. Even though this set of initial conditions is 
the most dynamic case of the cluster models with $r_{0.5}=0.8$~pc, only one cluster, 
with a mass of $10^{3.5}~\rmn{M}_{\sun}$, ejects its $\rmn{S_{MAXI}}$ (Table~3). 
But the initially second heaviest star of the cluster, 
which becomes the most massive one in the cluster after ejection of $\rmn{S_{MAXI}}$, 
has a similar mass to the mass of $\rmn{S_{MAXI}}$. Therefore the effect of the ejection on the 
$m_{\rmn{max}}$--$M_{\rmn{ecl}}$ relation is negligible in this case. 

It is unlikely that the $m_{\rmn{max}}$--$M_{\rmn{ecl}}$ relation is affected
by dynamical ejection of $\rmn{S_{MAXI}}$ for clusters of this size.
But it is worthy to note that at 3~Myr a few OP clusters show their $\rmn{S_{MAXI}}$ moving faster 
than the escape velocity, $v_{\rmn{esc}}(r) = \sqrt{2|\Phi(r)|}$, where $\Phi(r)$ is the 
gravitational potential at a distance $r$ from the cluster centre, 
although these clusters barely eject their $\rmn{S_{MAXI}}$ (Table~3). 

\subsubsection{The $r_{0.5} = 0.3$~pc models}
Fig.~3 shows the $m_{\rmn{max}}$--$M_{\rmn{ecl}}$ relation for the binary-rich cluster models 
with $r_{0.5}=0.3$~pc. Despite the smaller size of the clusters, 
none of the single star clusters (NMS3S and MS3S) eject their most massive star 
(Table~3). Only two massive clusters with $M_\rmn{ecl}=10^{3.5} \rmn{M}_{\sun}$ eject 
their $\rmn{S_{MAXI}}$ when the massive binaries are randomly paired (NMS3RP and MS3RP in Table~3). 
However binaries help the most massive star to attain a higher velocity compared 
to the single-star clusters. 
The number of clusters for which the heaviest star has a speed exceeding the escape velocity 
is larger when the stars are initially in a binary system (Table~3). 

Clusters with $M_{\rmn{ecl}}\leq10^{2.5}~\rmn{M}_{\sun}$ hardly eject
their most massive star even though the clusters form in energetic initial conditions 
such as being binary-rich and mass-segregated. 
Among all clusters with $M_{\rmn{ecl}}\leq10^{2.5}~\rmn{M}_{\sun}$,
only one out of 100 clusters with $M_{\rmn{ecl}}=10^2~\rmn{M}_{\sun}$ and two out of 100 clusters 
with $M_{\rmn{ecl}}=10^{2.5}~\rmn{M}_{\odot}$  
eject their $\rmn{S_{MAXI}}$ (Table~3). However at 3~Myr there are a few clusters whose 
$\rmn{S_{MAXI}}$ is inside the tidal radius but has a velocity higher than 
the escape velocity so that it may leave the cluster later.    

For clusters with $M_{\rmn{ecl}}\geq10^{3}~\rmn{M}_{\sun}$, 15-40 per cent of 
the most energetic models (MS3OP) have $\rmn{S_{MAXI}}$ moving faster than the escape velocity 
and 2-20 per cent of the clusters have ejected their $\rmn{S_{MAXI}}$ at 3~Myr. 
When the clusters are initially mass-segregated, slightly more clusters eject $\rmn{S_{MAXI}}$.

Fig.~4 shows the ejection frequency of $\rmn{S_{MAXI}}$ 
as a function of cluster mass for the MS3OP models. We only plot these models since
the other models barely eject their initially most massive star (see Table~3 for all models).
The ejection probability of the $\rmn{S_{MAXI}}$ star increases with the cluster mass 
as the stellar density increases. About 8 (20) per cent of $10^{3.5}$ ($10^{4}$)~$\rmn{M}_{\sun}$ 
clusters eject their $\rmn{S_{MAXI}}$ within 3~Myr (squares in Fig.~4).   
Some massive clusters which eject their $\rmn{S_{MAXI}}$ can be missed as we use the tidal radius 
as a criterion for the ejection. Some ejected $\rmn{S_{MAXI}}$ from massive clusters 
may not reach the clusters' tidal radius by 3~Myr due to their large tidal radius.
Thus, the real ejection frequency would be higher than the above value for the massive clusters. 
By using $N_{\rmn{vesc10}}$ (Table~3), the number of clusters whose $\rmn{S_{MAXI}}$ has a velocity 
greater than the escape velocity of the cluster and has travelled beyond 10~pc from the cluster centre,
the ejection frequency increases to 40 per cent for the cluster with $10^4~\rmn{M}_{\sun}$ (circles in Fig.~4).   
In order to estimate the real probability for the $\rmn{S_{MAXI}}$ ejection, 
in addition, we provide in Appendix~A the number of clusters whose $\rmn{S_{MAXI}}$ is found 
beyond the distance criterion using the half-mass radius for all models.

\begin{figure}
 \centering
 \includegraphics[width=80mm]{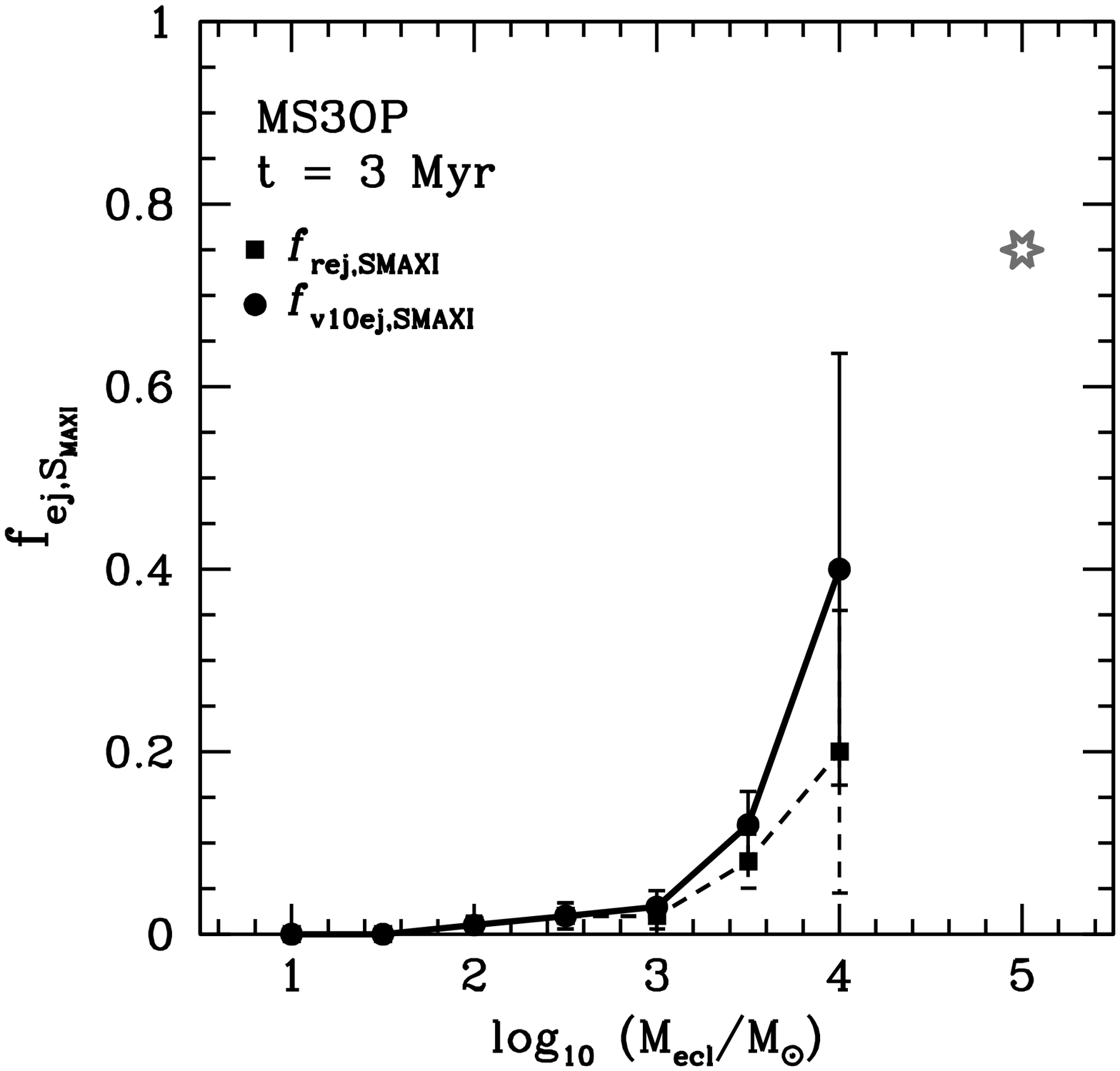}
 \caption{Ejection frequency of the $\rmn{S_{MAXI}}$ as a function of cluster mass for mass-segregated clusters 
 with massive binaries (MS3OP) at 3~Myr. $f_{\rmn{rej,S_{MAXI}}}$, marked with squares, is the ratio of
 the number of clusters whose $\rmn{S_{MAXI}}$ is located further than the tidal radius from
 the cluster centre, $N_{\rmn{resc}}$ in Table~3, to the total number of the realizations, $N_{\rmn{run}}$, 
 which is 100 in our study for each initial condition set (10 for $10^4~\rmn{M}_{\sun}$ clusters). 
 Errorbars indicate Poisson uncertainties. All other models have $f_{\rmn{rej,S_{MAXI}}} \approx 0$ (Table~3).
 The circles, $f_{\rmn{v10ej,S_{MAXI}}}$, are the ratio of $N_{\rmn{vesc10}}$ (Table~3) to $N_{\rmn{run}}$. 
 $f_{\rmn{v10ej,S_{MAXI}}}$ is probably closer to the real ejection frequency as some massive clusters 
 which eject their $\rmn{S_{MAXI}}$ can be missed in $N_{\rmn{resc}}$ due to their large tidal radii. 
 The grey star is the ejection frequency of $\rmn{S_{MAXI}}$ from clusters with $M_{\rmn{ecl}}=10^5~\rmn{M}_{\sun}$ 
 taken from the calculations by \citet{BKO12}. Those authors refer to a star that is ejected when 
 its distance from the cluster centre is larger than 10~pc. Note that the initial conditions 
 of their calculations are different but comparable to our MS3OP models. 
 }
\end{figure}

From Figs~2 and~3 it is evident that collisions of massive stars may conceal the dynamical ejection of 
$\rmn{S_{MAXI}}$ by the product of the collisions becoming more massive
than the mass of the initially heaviest member. Thus the dynamical behaviour of 
the $\rmn{S_{MAXI}}$ needs to be considered to distinguish whether 
it is ejected or not. Figs~5 and ~6 show the distance from the cluster centre 
and the velocity of the $\rmn{S_{MAXI}}$ in 
NMS3OP and MS3OP clusters with $10~\rmn{M}_{\sun}$ and $10^{3.5}~\rmn{M}_{\sun}$. 

The heaviest stars are initially located at a wide range of radii up to $\approx$ 2.5~pc 
in the unsegregated cluster models (Fig.~5) 
while they are centrally concentrated in the segregated ones (Fig.~6). 
In both cases, massive stars sink towards the centre of the clusters due to dynamical friction 
and/or energy equipartition. This is more prominent in the unsegregated clusters 
since the heaviest stars already reside in the deep potential of the segregated clusters 
at the beginning of the calculations. 

As shown in Figs~5 and~6, the heaviest stars hardly attain a high velocity
if the massive stars are randomly paired into binaries. 
In random pairing, massive stars are mostly paired with low mass stars therefore
their mass ratios ($m_{2}/m_{1}$, where $m_{1}\geq m_{2}$) are skewed to 0.
Randomly paired massive binaries therefore behave like single stars. 
Clusters with massive binaries paired randomly do not shoot out their 
$\rmn{S_{MAXI}}$ more frequently even though the clusters are initially mass-segregated.

Clusters with massive binaries paired by OP and $M_{\rmn{ecl}} \geq 100~\rmn{M}_{\sun}$ 
effectively produce heaviest stars with velocities exceeding the escape velocity of the cluster.
For example, in the case of some initial conditions more than 20 per cent of the clusters show 
that their $\rmn{S_{MAXI}}$ has a velocity larger than the escape velocity at 3~Myr (Table~3).
It is known that massive stars are ejected from the small core of massive stars in the 
cluster centre lacking low mass stars \citep{CP92}. In the case of segregated clusters 
with OP massive binaries, the clusters already form with this kind of core 
thus the massive stars can be ejected at a very early age of the cluster. 
$S_\rmn{MAXI}$ of one cluster with $M_{\rmn{ecl}}=10^{3.5}~M_{\sun}$ from the MS3OP model
has travelled more than 200~pc from the cluster with a velocity of about $80~\rmn{km~s}^{-1}$ 
at 3~Myr (Fig.~6).

Although the low mass clusters barely eject their $\rmn{S_{MAXI}}$ regardless of 
their initial conditions, the binary-rich or binary-poor cases show differences, 
e.g. the number of clusters whose $\rmn{S_{MAXI}}$ has a velocity 
greater than the escape velocity is larger for binary-rich clusters.
For massive clusters, on the contrary, binary-poor and binary-rich clusters with random pairing 
show similar results in the case of unsegregated models.
Primordial mass segregation enhances the ejection of the most massive star at earlier times 
and helps more clusters shoot out their $\rmn{S_{MAXI}}$ within 3~Myr.

\begin{figure*}
 \centering
 \includegraphics[width=75mm]{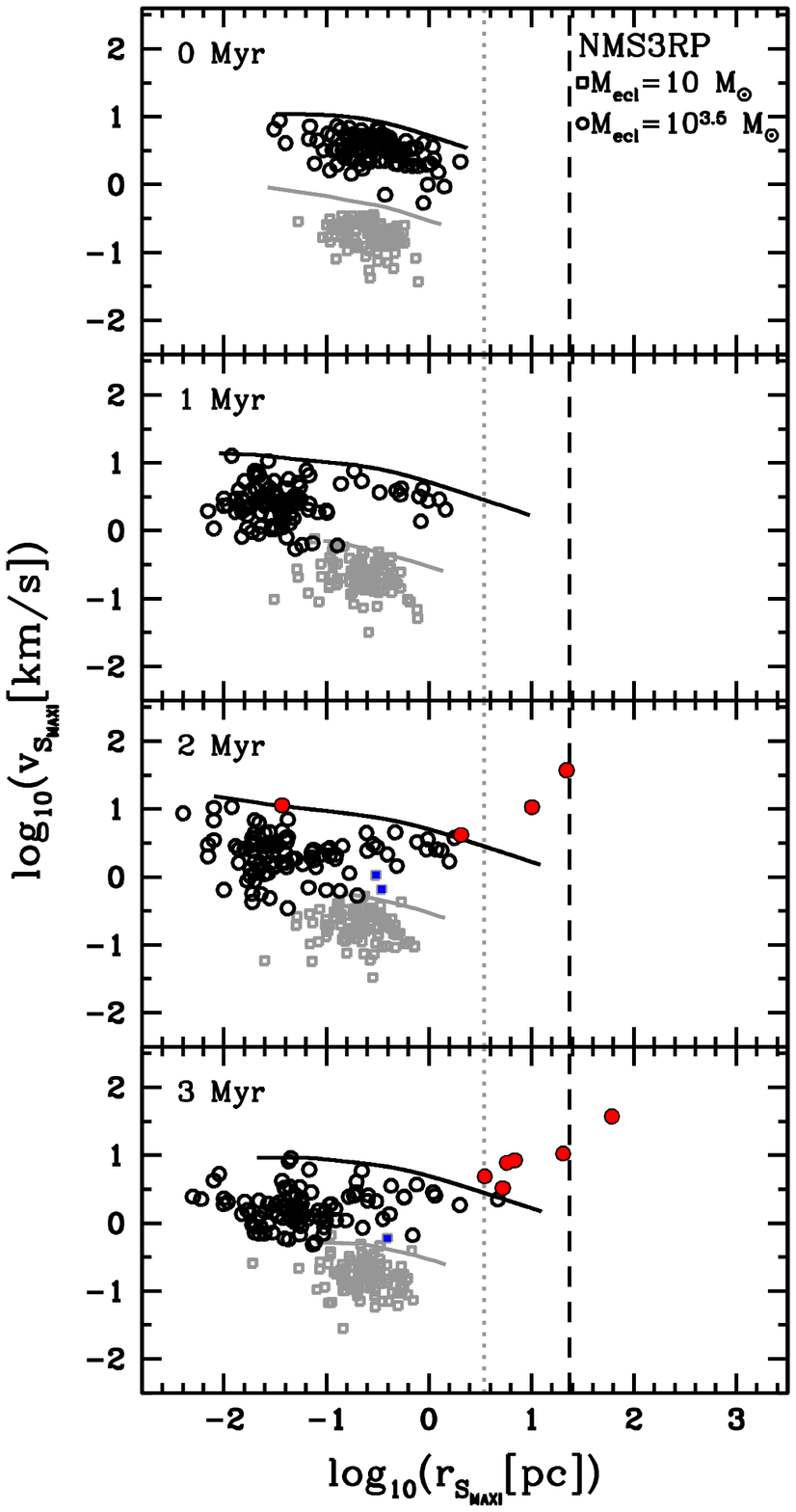}
 \includegraphics[width=75mm]{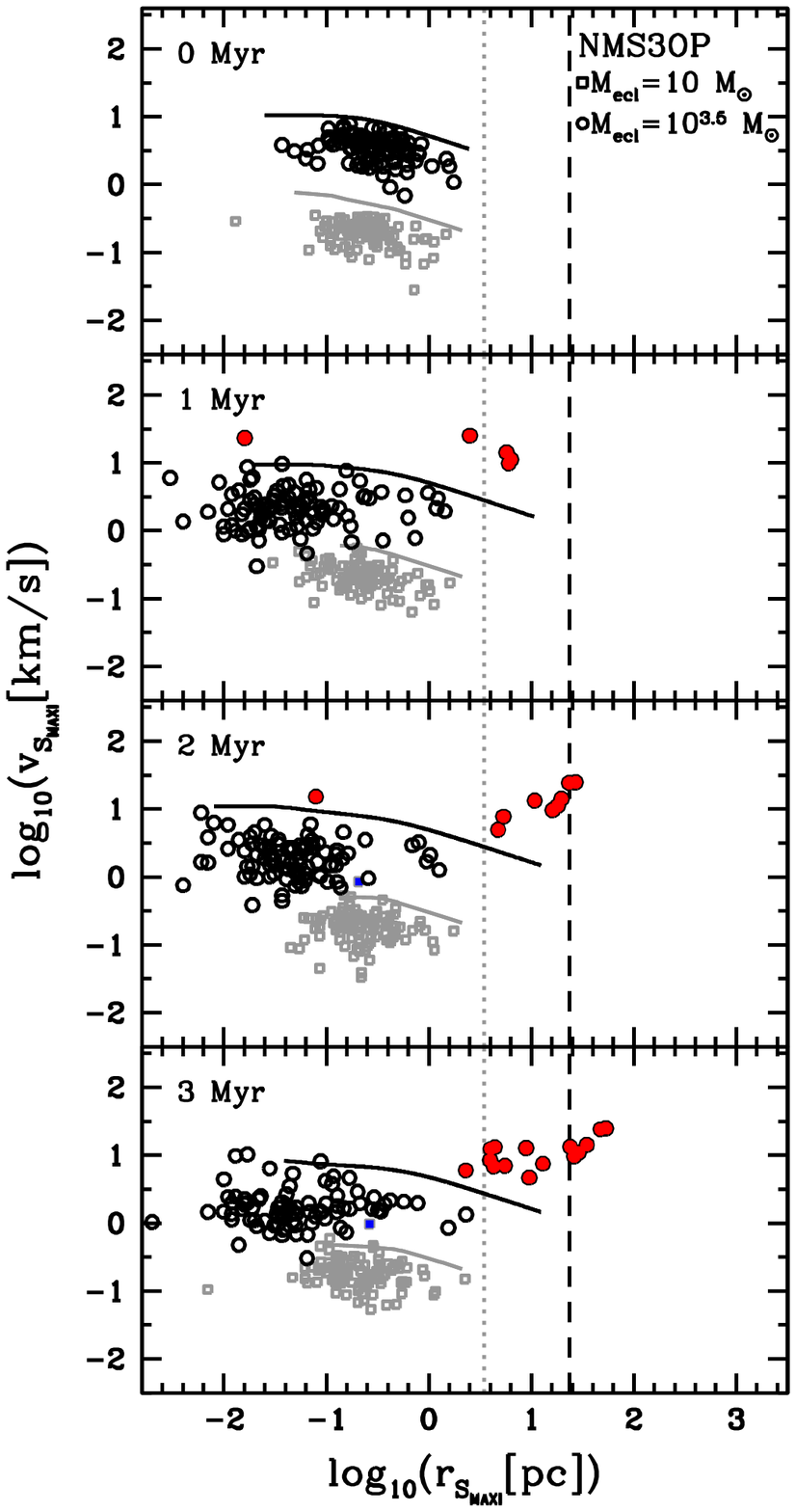}
 \caption{ Distances from the cluster centre and velocities of $\rmn{S_{MAXI}}$
 stars of the initially unsegregated binary-rich clusters with $r_{0.5}=0.3$~pc
 ({\it left}: NMS3RP, {\it right}: NMS3OP).
 Each dot denotes a cluster and in total there are 100 dots per configuration. 
 Squares and circles are the values of $\rmn{S_{MAXI}}$
 in the clusters with $10~\rmn{M}_{\sun}$ and $10^{3.5}~\rmn{M}_{\sun}$, respectively.
 Filled (red) symbols represent $\rmn{S_{MAXI}}$ stars with a speed exceeding the escape velocity.
 The black (grey) solid curves are the escape velocity of one cluster with $10^{3.5}$ 
 ($10$)~$\rmn{M}_{\sun}$ as a function of distance from the cluster centre at each Myr.
 The grey dotted and the black dashed vertical lines indicate the initial tidal radius
 of the cluster with $10~\rmn{M}_{\sun}$ and with $10^{3.5}~\rmn{M}_{\sun}$, respectively.
 Note how mass segregation develops by 1~Myr and how OP increases the occurrence of ejected most massive stars.}
\end{figure*}

\begin{figure*}
 \centering
 \includegraphics[width=75mm]{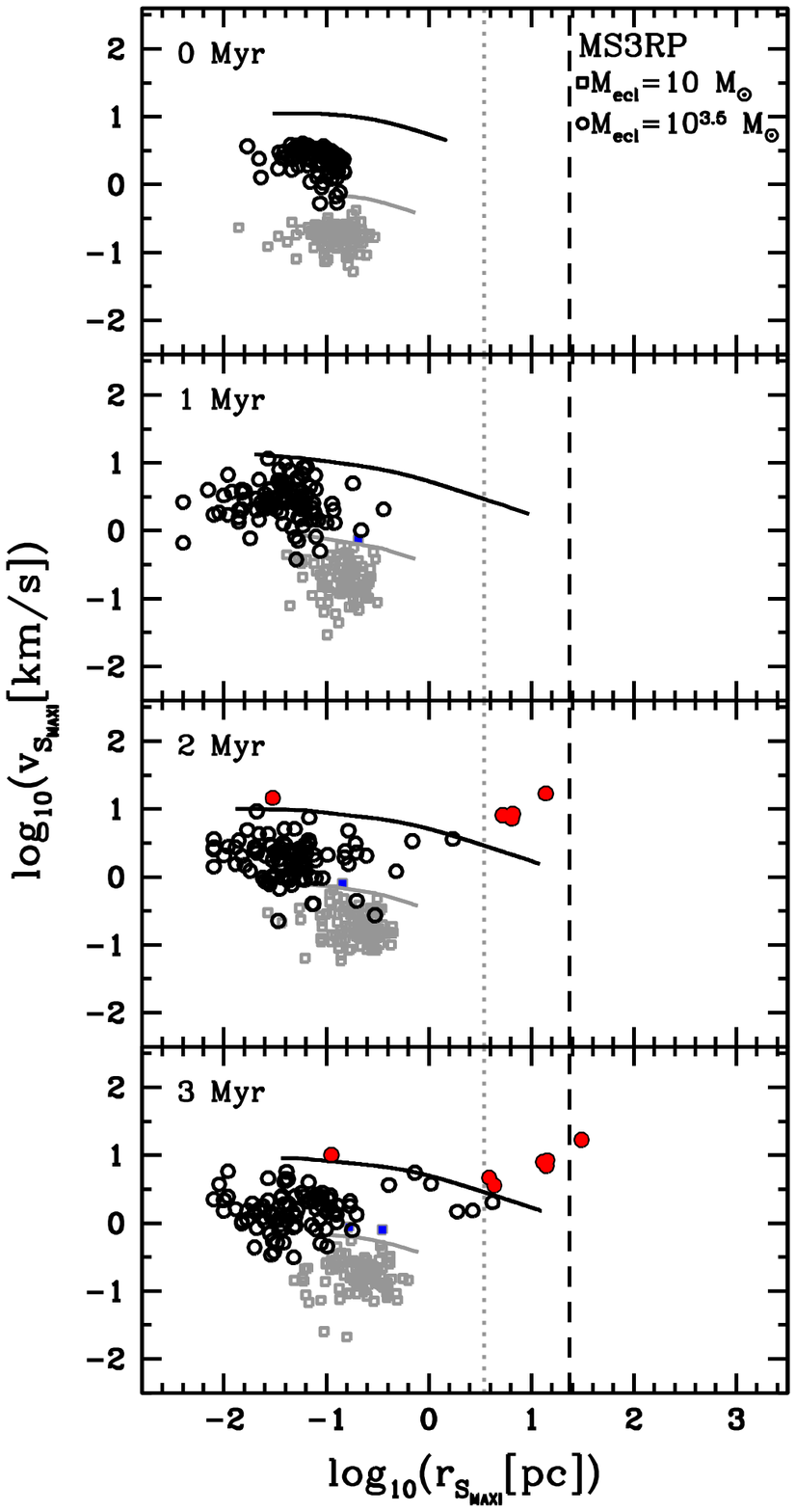}
 \includegraphics[width=75mm]{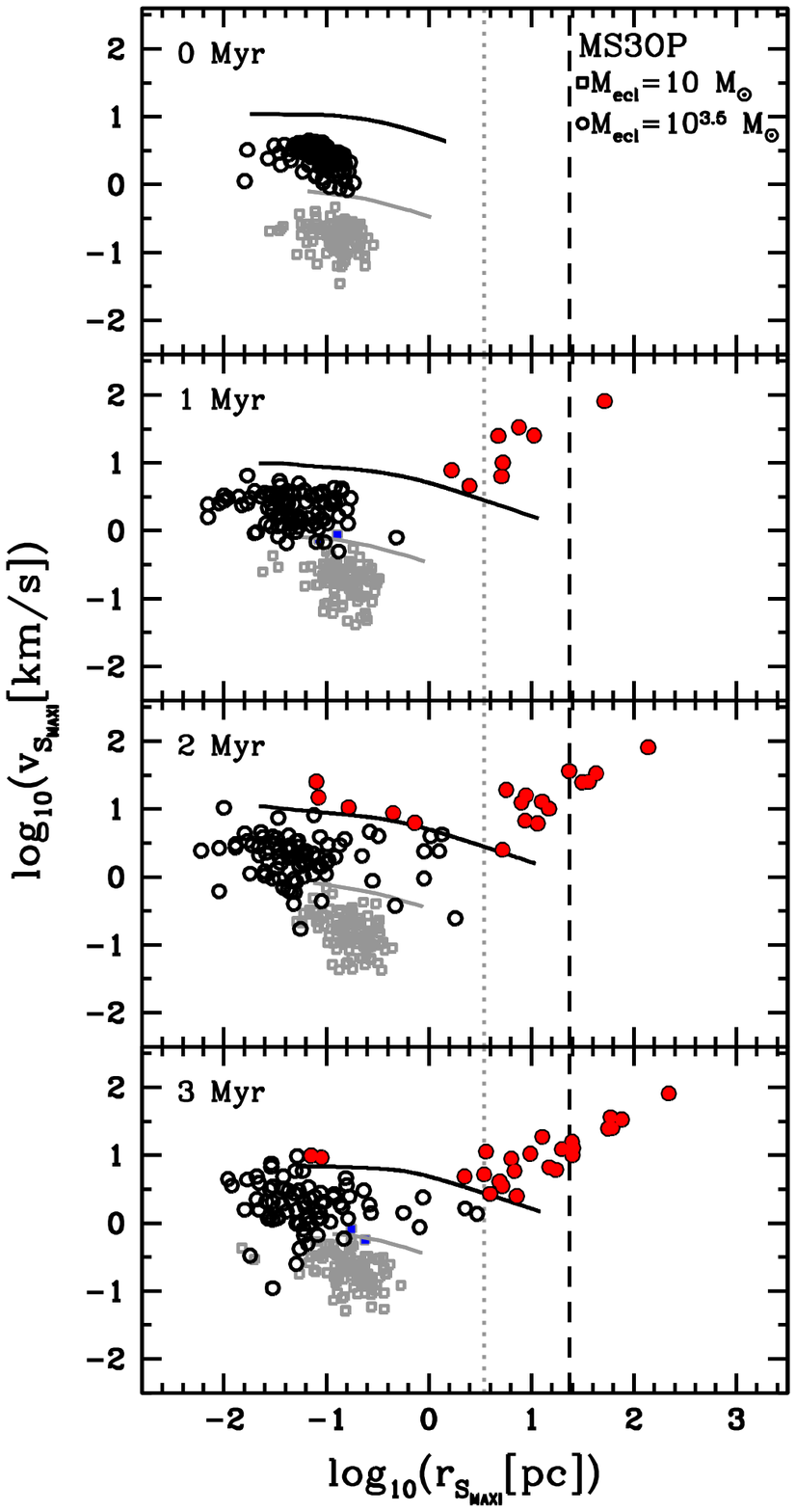}
 \caption{Same as Fig.~5 but for the initially mass-segregated binary-rich clusters
 with $r_{0.5} = 0.3$~pc ({\it left}: MS3RP, {\it right}: MS3OP). }
\end{figure*}

\begin{table*}
 \centering
 \caption{
 Results at 3~Myr. $M_{\rmn{ecl}}$ is a cluster mass in $\rmn{M_{\sun}}$ 
 (per $M_{\rmn{ecl}}$ value there are 100 clusters, but 10 clusters for $10^4~\rmn{M}_{\sun}$).
 $N_{\rmn{resc}}$ is the number of clusters whose $\rmn{S_{MAXI}}$ is located beyond 
 the tidal radius of the cluster. $N_{\rmn{vesc}}$, the number of clusters whose 
 $\rmn{S_{MAXI}}$ has a velocity larger than the escape velocity. 
 $N_{\rmn{vesc10}}$, the number of clusters whose $\rmn{S_{MAXI}}$ is located further 
 than 10~pc from the cluster centre and has a velocity larger than the escape velocity. 
 $N_{\rmn{c}}$ is the number of clusters whose $m_{\rmn{max}}$ changes due to stellar 
 collisions. Numbers in brackets indicate the collision products that do not involve $\rmn{S_{MAXI}}$. 
 For example, in the case of clusters with $M_{\rmn{ecl}}=10^{3.5}~\rmn{M}_{\sun}$ 
 from the NMS3RP model, 1 out of 100 clusters lost their $\rmn{S_{MAXI}}$ by dynamical ejection, 
 $\rmn{S_{MAXI}}$ of 6 clusters have a velocity greater than the escape velocity of the cluster, 
 for 2 clusters out of these 6 clusters the star is located beyond 10~pc from the cluster centre.
 Stellar collisions which change the $m_{\rmn{max}}$ have occurred in 51 clusters, 
 for 11 clusters out of these 51 the collisions do not involve $\rmn{S_{MAXI}}$.
 }
 \begin{tabular}{@{}lccccccccc@{}}
  \hline
 $M_{\rmn{ecl}} [\rmn{M_{\sun}}]$ & $N_{\rmn{resc}}$ & $N_{\rmn{vesc}}$ &
   $N_{\rmn{vesc10}}$  & $N_{\rmn{c}}$
 & & $N_{\rmn{resc}}$ & $N_{\rmn{vesc}}$ & $N_{\rmn{vesc10}}$ & $N_{\rmn{c}}$ \\
  \hline
 & \multicolumn{4}{c}{Unsegregated cluster model} & &
   \multicolumn{4}{c}{Mass-segregated cluster model} \\
 & \multicolumn{4}{l}{NMS3S} & & \multicolumn{4}{l}{MS3S} \\
 10         & 0 & 0 & 0 & 0 & & 0 & 0 & 0 & 0 \\
 10$^{1.5}$ & 0 & 0 & 0 & 0 & & 0 & 0 & 0 & 0 \\
 10$^{2}$   & 0 & 0 & 0 & 0 & & 0 & 0 & 0 & 0 \\
 10$^{2.5}$ & 0 & 0 & 0 & 0 & & 0 & 1 & 0 & 2 \\
 10$^{3}$   & 0 & 0 & 0 & 6 (1) & & 0 & 2 & 0 & 8 \\
 10$^{3.5}$ & 0 & 6 & 1 & 36 (8)& & 0 & 5 & 1 & 36 (4) \\
 & \multicolumn{4}{l}{NMS3RP} & & \multicolumn{4}{l}{MS3RP} \\
 10         & 0 & 1 & 0 & 0 & & 0 & 2 & 0 & 0 \\
 10$^{1.5}$ & 0 & 0 & 0 & 0 & & 0 & 2 & 0 & 1 (1) \\
 10$^{2}$   & 0 & 1 & 0 & 1 & & 0 & 0 & 0 & 2 \\
 10$^{2.5}$ & 0 & 0 & 0 & 8 (2) & & 0 & 4 & 0 & 9 (1) \\
 10$^{3}$   & 0 & 2 & 0 & 17 (3) & & 1 & 4 & 1 & 23 (1) \\
 10$^{3.5}$ & 1 & 6 & 2 & 51 (11) & & 1 & 7 & 4 & 46 (4)\\
 & \multicolumn{4}{l}{NMS3OP} & & \multicolumn{4}{l}{MS3OP} \\
 10         & 0 & 1  & 0 & 0 & & 0 & 2  & 0 & 0 \\
 10$^{1.5}$ & 0 & 4  & 0 & 2 & & 0 & 2  & 0 & 0 \\
 10$^{2}$   & 0 & 5  & 0 & 1 & & 1 & 5  & 1 & 1 \\
 10$^{2.5}$ & 0 & 14 & 0 & 5 (1) & & 2 & 23 & 2 & 4 \\
 10$^{3}$   & 1 & 15 & 2 & 23 (4) & & 2 & 15 & 3 & 17 (3)\\
 10$^{3.5}$ & 6 & 15 & 7 & 60 (23) & & 8 & 24 & 12 & 56 (14)\\
 10$^{4}$   &   &    &   &         & & 2 & 4 & 4 & 9 (4)\\
 & \multicolumn{4}{l}{NMS8S} & & \multicolumn{4}{l}{MS8S} \\
 10         & 1 & 0 & 0 & 0 & & 0 & 0 & 0 & 0 \\
 10$^{1.5}$ & 0 & 0 & 0 & 0 & & 0 & 0 & 0 & 0 \\
 10$^{2}$   & 0 & 0 & 0 & 0 & & 0 & 0 & 0 & 0 \\
 10$^{2.5}$ & 0 & 0 & 0 & 0 & & 0 & 0 & 0 & 0 \\
 10$^{3}$   & 0 & 0 & 0 & 0 & & 0 & 0 & 0 & 0 \\
 10$^{3.5}$ & 0 & 0 & 0 & 1 & & 0 & 0 & 0 & 0 \\
 & \multicolumn{4}{l}{NMS8RP} & & \multicolumn{4}{l}{MS8RP} \\
 10         & 0 & 1 & 0 & 0 & & 0 & 1 & 0 & 0 \\
 10$^{1.5}$ & 0 & 1 & 0 & 0 & & 0 & 0 & 0 & 0 \\
 10$^{2}$   & 0 & 0 & 0 & 1 & & 0 & 0 & 0 & 0 \\
 10$^{2.5}$ & 0 & 0 & 0 & 1 & & 0 & 0 & 0 & 1 \\
 10$^{3}$   & 0 & 0 & 0 & 2 (1) & & 0 & 1 & 0 & 0 \\
 10$^{3.5}$ & 0 & 0 & 0 & 5 & & 0 & 0 & 0 & 1 (1) \\
 & \multicolumn{4}{l}{NMS8OP} & & \multicolumn{4}{l}{MS8OP}  \\
 10         & 0 & 3 & 0 & 0 & & 0 & 0 & 0 & 0 \\
 10$^{1.5}$ & 0 & 1 & 0 & 1 & & 0 & 1 & 0 & 0 \\
 10$^{2}$   & 0 & 1 & 0 & 1 (1) & & 0 & 0 & 0 & 0 \\
 10$^{2.5}$ & 0 & 0 & 0 & 0 & & 0 & 3 & 0 & 0 \\
 10$^{3}$   & 0 & 2 & 0 & 2 & & 0 & 5 & 0 & 5 (1)\\
 10$^{3.5}$ & 0 & 3 & 0 & 10 (4) & & 1 & 3 & 1 & 8 (6)\\
 \hline
 \end{tabular}
\end{table*}

\subsection {Stellar collisions}
$m_{\rmn{max}}$ can increase by stellar collisions, either involving $\rmn{S_{MAXI}}$ or not,
which makes a star heavier than the mass of $\rmn{S_{MAXI}}$ before the collisions occur. 
Although stellar collisions occur over a whole range of stellar masses, in this study, 
we only care about the collision that changes $m_{\rmn{max}}$. 
The occurrences of the collisions are contained in Table~3.

For clusters with $r_{0.5}=0.8$~pc 
($2~\rmn{M}_{\sun}~\rmn{pc}^{-3} < \rho_{0.5} < 750~\rmn{M}_{\sun}~\rmn{pc}^{-3}$, where 
$\rho_{0.5}=3M_{\rmn{ecl}}/(8 \pi r_{0.5}^3)$ is the average mass density within the half-mass radius)  
stellar collisions rarely occur in them due to their low density.
In single-star cluster models (NMS8S and MS8S), only one cluster shows that 
its $m_{\rmn{max}}$ changes by a stellar collision which involves $\rmn{S_{MAXI}}$.
In the case of binary-rich models there are a few clusters whose $m_{\rmn{max}}$ changes via stellar 
collisions, mostly in the most massive cluster models, but with a probability of less than 10 per cent 
taking all models into account.
In low-density ($r_{0.5}=0.8$~pc) clusters a change of $m_{\rmn{max}}$ through 
stellar collisions is highly improbable.

For clusters with $r_{0.5}=0.3$~pc, the stellar collision result of low-mass clusters, 
$M_{\rmn{ecl}}\leq10^{2.5}~\rmn{M}_{\sun}$ ($\rho_{0.5} \lesssim 1400~\rmn{M}_{\sun}~\rmn{pc}^{-3}$) 
is similar to the result from clusters with $r_{0.5}=0.8$~pc. But in the case of the 
massive cluster models ($M_{\rmn{ecl}}\geq10^3~\rmn{M}_{\sun}$), especially the most massive one, 
about half of them experience a change of $m_{\rmn{max}}$ by stellar collisions.

Most of the collisions are induced by binary-encounters, in which
the two components of a binary system collide due to their highly eccentric orbit generated
by perturbation through other stars. Direct dynamical collisions are extremely rare.
This naturally explains why $N_{\rmn{c}}$ in Table~3 becomes larger when clusters are initially
binary-rich.
And the number increases when massive binaries are paired with the OP method when compared to RP.
This can be understood because both components of a massive binary in the OP method
are massive stars and thus have larger sizes leading to collisions.
With a different collision channel from our result, Gaburov, Gualandris \& Portegies Zwart (2008) 
also showed that a binary is more efficient in stellar collisions than a single star as the 
enhanced cross-section of a binary compared to a single star results in other stars engaging 
the binary, and then this can lead to a collision between one of the binary components and the incoming third star. 

Both the dynamical ejection of the $\rmn{S_{MAXI}}$ and stellar collisions that
increase $m_{\rmn{max}}$ barely take place in the same cluster within the cluster mass range we study.
As shown by \citet{BK11} and \citet{MC11}, stellar collisions generally lead to
the formation of a single very massive star through the merging of several massive stars
rather than the formation of many massive stars. This single very massive star is hardly
ejected from the cluster and its formation reduces the number of massive
stars thus it may hamper the ejection of other massive stars.
Out of 7200 cluster models (excluding the $10^4~\rmn{M}_{\sun}$ clusters from MS3OP model), 
there are only three clusters in which both events occur.  
In one cluster $\rmn{S_{MAXI}}$ is dynamically ejected, but one massive binary in the cluster merges then 
becoming more massive than $\rmn{S_{MAXI}}$. 
In the other two clusters their $\rmn{S_{MAXI}}$ gains mass by the collision with another star 
and then it is dynamically ejected. 

Although this event is very rare, the most massive star in the star-forming region
LH~95 in Large Magellanic Cloud \citep{Ret12} might be an example. 
The peculiarities of the star, with a much younger age than the average age of other stars in the region 
and the mass being higher than $m_{\rmn{max}}$ from the \citet{WK04} $m_{\rmn{max}}$--$M_{\rmn{ecl}}$ relation, 
could have resulted from a stellar collision between binary components with similar masses 
of which the primary star may have been the initially most massive star of one of the three main substructures in the region, 
with a mass following the $m_{\rmn{max}}$--$M_{\rmn{ecl}}$ relation from \citet{WK04}. 
The stellar collision could have increased the stellar mass and rejuvenated the star. 
And the ejection of the star with a low velocity can explain the location of the star which is at a rather 
far ($\approx10$~pc) distance from any of the substructures but still within the region.

\subsection{The spread of $m_{\rmn{max}}$}
\begin{figure}
 \centering
 \includegraphics[width=85mm]{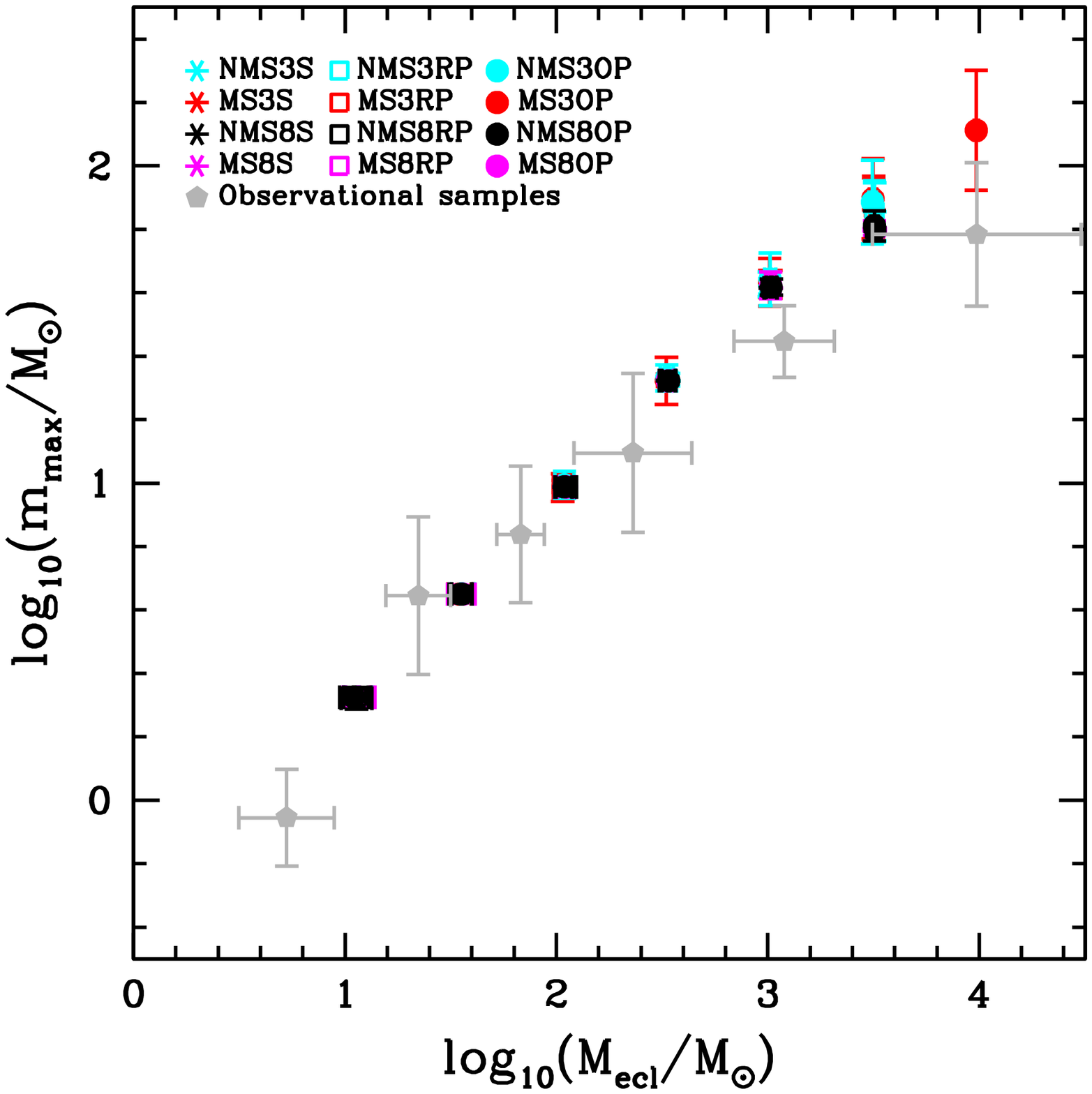}\\
 \includegraphics[width=85mm]{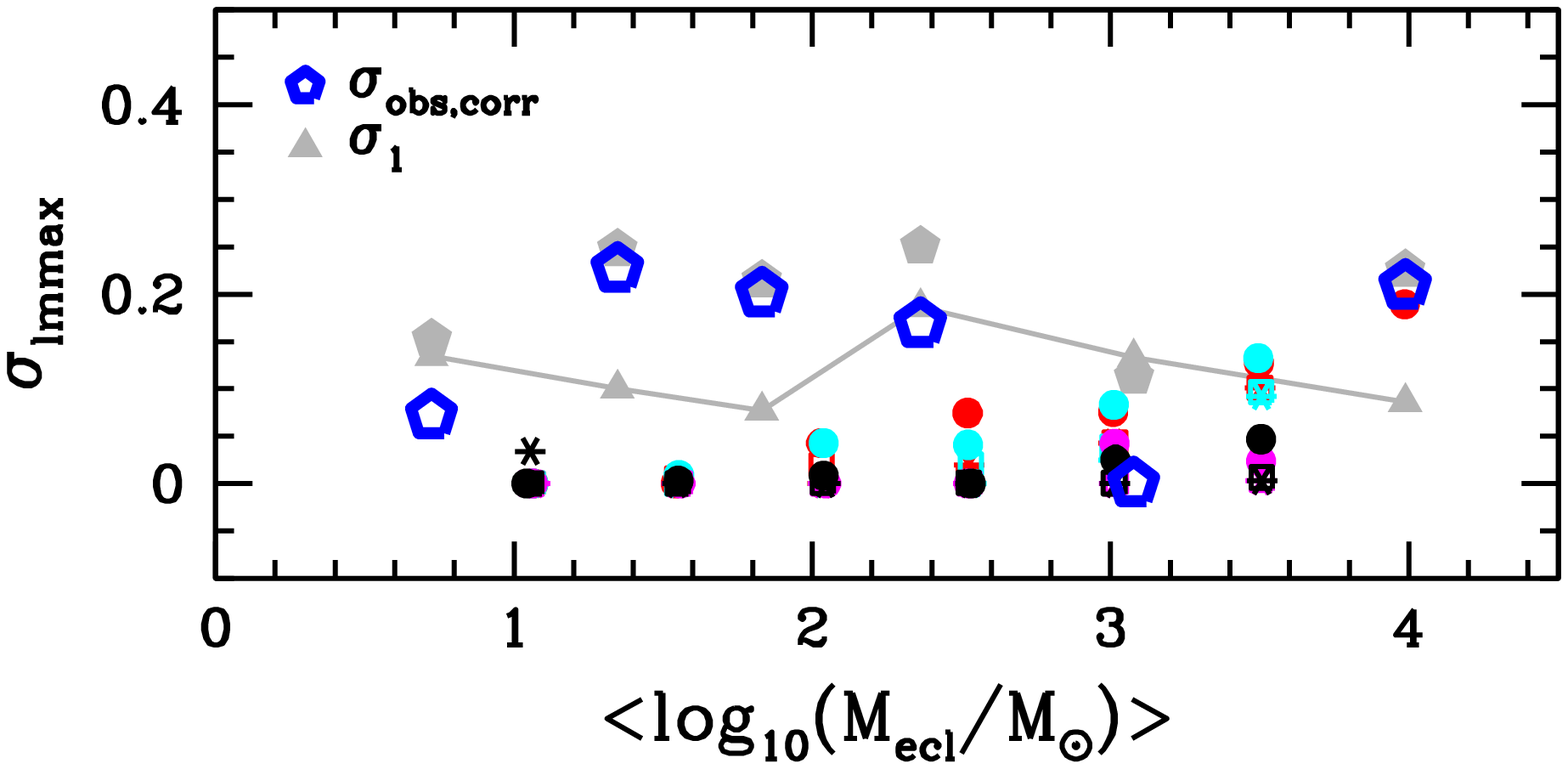}\\
 \caption{Top: Average $m_{\rmn{max}}-M_{\rmn{ecl}}$ plot for all models at 3~Myr. 
 Errorbars indicate the standard deviations of $M_{\rmn{ecl}}$ and $m_{\rmn{max}}$. 
 Grey symbols are the observed cluster samples as in Figs~2-~3. 
 Colour and symbol codes are given in the upper-left corner of the figure. 
 Bottom: Standard deviation of $\log_{10} m_{\rmn{max}}$, $\sigma_{\rmn{lmmax}}$, 
 for each cluster model from the top figure. Colours and symbols are the 
 same as in the top panel. Note that for $M_{\rmn{ecl}}\leq10^{2.5}~\rmn{M}_{\sun}$ 
 the observed $\sigma_{\rmn{lmmax}}$, $\sigma_{\rmn{obs}}$, is much larger than the 
 dispersion in the models. To quantify the deviation due to binning the cluster mass in the observed 
 range of cluster masses in each mass bin, 
 we obtain the $m_{\rmn{max}}$ according to equations~1 and~2 with 
 $m_{\rmn{max}}^{*}=150~\rmn{M}_{\sun}$ for the observed cluster masses. 
 We then calculate the standard deviation of $\log_{10} m_{\rmn{max}}$, $\sigma_{1}$, 
 by binning the clusters in the same way as for $\sigma_{\rmn{obs}}$.
 This is shown as grey triangles connected with a solid line. 
 The corrected observed $\sigma_{\rmn{lmmax}}$, $\sigma_{\rmn{obs,corr}} = \sqrt{\sigma_{\rmn{obs}}^2-\sigma_{1}^2}$  
 (if $\sigma_{\rmn{obs}}>\sigma_{1}$ otherwise $\sigma_{\rmn{obs,corr}}=0$), 
 is plotted with blue open pentagons.
  }
\end{figure}

Fig.~7 presents the standard deviation of $\log m_{\rmn{max}}$, $\sigma_{\rmn{lmmax}}$, 
for the observed samples and our models. The observed larger $m_{\rmn{max}}$ spread 
than what emerges from our models may be a result of stochastic effects of star formation 
as the dynamical processes hardly influence the change of $m_{\rmn{max}}$ in these clusters, 
as shown by this study. However, numerical simulations of star cluster formation show 
that $m_{\rmn{max}}$ and $M_{\rmn{ecl}}$ evolve tightly following the relation 
of $m_{\rmn{max}} \propto M_{\rmn{ecl}}^{2/3}$ \citep{BVB04,Pet10}.  

For low-mass clusters ($M_{\rmn{ecl}} < 100~\rmn{M}_{\sun}$) the differences between 
$\sigma_{\rmn{lmmax}}$ of the observation and our model are large despite taking into account 
the spread due to the different cluster masses in the bins of observational data. 
This could be due to the large uncertainties of the observations. Note that the homogeneous data set 
from \citet{KM11} is more confined to the relation than the inhomogeneous data from 
\citet{WKB10} which come from many different references.

The spread in our models for relatively massive clusters ($M_\rmn{ecl}\geq10^3~\rmn{M}_{\sun}$) 
is comparable to the observed one in the case of clusters with $r_{0.5}=0.3$~pc and with OP for
massive binaries (NMS3OP and MS3OP models).

Stellar collisions dominate the change of $m_{\rmn{max}}$ of relatively dense clusters 
with $M_{\rmn{ecl}}\geq 10^{2.5}~\rmn{M}_{\sun}$ and $r_{0.5}=0.3$~pc (Table~3) 
and therefore the spread of $m_{\rmn{max}}$ of these clusters mostly comes from the collisions.
However it should be noted that the exact solution for the stellar collision process
is as yet poorly understood. The treatment of stellar collisions in the code is simply 
adding the masses of two stars. Thus the collision products shown in this study provide 
only a rough idea about merger rates and their masses. 
Therefore we cannot quantify the real spread produced by the collisions
but we expect that it would be smaller than our result since in reality not all
the stellar mass ends up being in the merger product.

\section{Discussion}

\begin{figure}
 \centering
 \includegraphics[width=85mm]{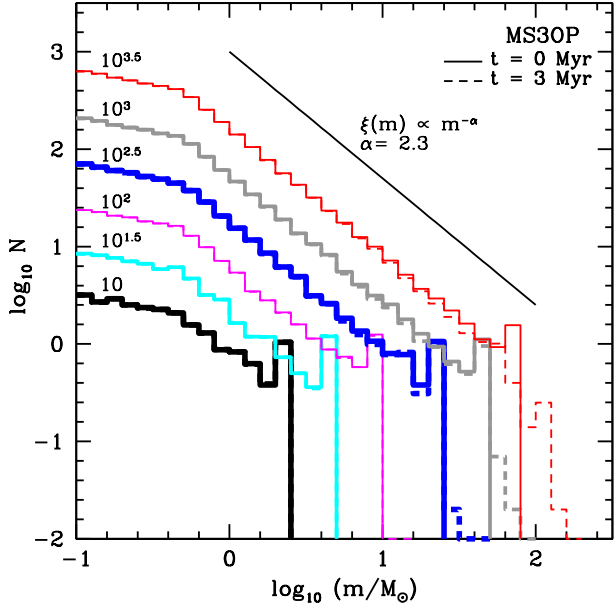}
 \caption{Mass functions of MS3OP clusters. Each line indicates the average mass function of
 clusters obtained from the 100 computations per models with a mass of $10$, $10^{1.5}$, 
 $10^2$, $10^{2.5}$, $10^3$ and $10^{3.5}~\rmn{M}_{\sun}$ from the bottom to the top. 
 The solid and dashed histograms are mass functions at 0 (initial) and 3~Myr,
 respectively. A solid line on top of the histograms is the Salpeter mass function. }
\end{figure}

Fig.~8 shows the initial- and the final ($t=3$~Myr) mass functions of MS3OP model clusters.
Massive stars are overpopulated in a cluster with a bump at the most massive stellar-mass-bin of 
the cluster's mass function because we enforce all clusters to have a star with a mass of $m_{\rmn{max,WK}}$. 
As a necessary condition for massive star ejections is a small core of massive
stars, the overpopulation of massive stars would enhance the ejection of massive stars.
The other effect that might result from this choice is the over-spread of $m_{\rmn{max}}$ in 
low-mass (i.e. small-number) clusters if many of the clusters eject their $\rmn{S_{MAXI}}$. 
When stellar masses are randomly drawn from the IMF low-mass clusters would hardly have a star 
with a mass close to $m_{\rmn{max}}$ while massive clusters, that populate stars over 
the whole mass range, would have a few. 
However, the dynamical ejection of the most massive star from the low-mass cluster is 
highly improbable thus enforcing clusters to have a star with a $m_{\rmn{max,WK}}$ 
would not have an impact on the spread.

The orbital parameters of the massive binaries are important for the ejection of massive stars. 
But, our knowledge of their initial distribution functions is still poor. 
We use the same period distribution for massive binaries as that of low-mass stars. 
However, a high fraction of short-period massive binaries is suggested by observations 
\citep{SE11} implying that massive binaries may have a different initial period distribution, 
compared to that of low-mass binaries. 
Dynamical ejections of the most massive star might become more likely than presented here.  
Future work taking into account the most recent constraints on the period distribution of 
massive star-binaries will investigate this issue. 
The results for $M_{\rmn{ecl}}\leq10^3~\rmn{M}_{\sun}$ clusters presented here will, however, 
not be affected as such clusters do not contain many, if any, massive stars. 
 
Apart from our MS3OP sequence of models, our young star cluster library does not contain 
clusters more massive than $10^4~\rmn{M}_{\sun}$ as they are computationally expensive due 
to the large binary population, while the observational sample contains a few clusters 
more massive than $10^{4}~\rmn{M}_{\sun}$. 
At a higher density massive clusters may show the ejection of the heaviest star with
a higher probability. Indeed, 20 per cent of the MS3OP clusters with 
$M_{\rmn{ecl}}=10^4~\rmn{M}_{\sun}$ eject their $\rmn{S_{MAXI}}$ (Fig.~4).
Furthermore, 3 out of 4 clusters with $M_{\rmn{ecl}}=10^{5}~\rmn{M}_{\sun}$ computed 
by \citet{BKO12} eject their $\rmn{S_{MAXI}}$. 
Further work including such massive clusters will be carried out to study this issue.

\section{Conclusions}
We have established a large theoretical young  
star cluster library using Aarseth's direct N-body integration code, {\sevensize NBODY}6, 
with 72 different combinations of initial conditions such as cluster size and mass, 
initial binary population, and primordial mass segregation.
The library contains two different sizes of clusters, $r_{0.5}=0.3$ and $0.8$~pc, 
with two different binary fractions, zero and unity.  
In order to take into account that observed OB binaries likely have companions with similar masses,
we generate clusters with massive binaries not only paired randomly from the IMF but also having
a mass-ratio close to unity. In both cases the stars follow the canonical IMF.
And we model initially mass-segregated clusters as well as unsegregated clusters.
Using this library we study the $m_{\rmn{max}}$--$M_{\rmn{ecl}}$ relation during 
the early evolution ($\leq$~3~Myr) of the star clusters focusing on the effects on it established 
through dynamical ejection of the heaviest star from the cluster under the various 
initial conditions. Such computations of fully mass-segregated binary-rich clusters have never been 
performed before.

Stellar evolution, stellar collision and dynamical ejection can alter the $m_{\rmn{max}}$ 
of the cluster. In our models, all three effects are observed to affect $m_{\rmn{max}}$.
Stellar evolution affects the relation only for the massive ($\gtrsim10^{3}~\rmn{M}_{\sun}$) 
clusters since only stars more massive than 40~$\rmn{M}_{\sun}$ significantly lose 
their mass within 3~Myr. Furthermore, since the mass of the $\rmn{S_{MAXI}}$ is the same for 
the same cluster mass, stellar evolution cannot contribute to the spread of the $m_{\rmn{max}}$ 
in this study. Stellar collisions influence mainly massive clusters, especially 
for those with the smaller radius as the clusters are denser. For clusters with 
$r_{0.5}=0.3$~pc and $M_{\rmn{ecl}} = 10^{3.5}~\rmn{M}_{\sun}$, for about half of the clusters 
their $m_{\rmn{max}}$ has changed by stellar collisions within 3~Myr. 
Lastly, concerning the focus of this study, the dynamical ejection of $\rmn{S_{MAXI}}$ 
only occurs in the binary-rich clusters with $r_{0.5}=0.3$~pc and 
$M_{\rmn{ecl}}>10^{2}~\rmn{M_{\sun}}$. The number of the ejections increase when 
massive binaries are paired with the OP method and/or are initially concentrated 
in the cluster centre. As massive clusters likely have a few stars with masses close to the mass of $\rmn{S_{MAXI}}$, 
$m_{\rmn{max}}$ value would change little for massive clusters when $\rmn{S_{MAXI}}$ is dynamically ejected.
Overall we conclude that (dynamical) evolutionary effects 
hardly produce a spread of $m_{\rmn{max}}$ in the relation for low-mass 
($M_{\rmn{ecl}}\lesssim10^{2.5}~\rmn{M_{\sun}}$) or less dense ($r_{0.5}=0.8$~pc) clusters. 
Massive ($M_{\rmn{ecl}}=10^{3.5}~\rmn{M_{\sun}}$), binary-rich clusters with 
$r_{0.5}=0.3$~pc do show  a significant spread of $m_{\rmn{max}}$, mostly produced by stellar collisions, 
becoming comparable to the observed spread (Fig.~7).

Concerning the dynamical ejection of $\rmn{S_{MAXI}}$, we find that in general 
it is very unlikely even in relatively dense clusters with massive binaries with similar 
mass companions. However, its probability can reach up to 20 per cent in our cluster 
library depending on the initial configuration of the cluster. For example, none of the 
clusters without initial mass segregation but with OP binaries and with 
$M_{\rmn{ecl}}=10^{3.5}~\rmn{M_{\sun}}$ and $r_{0.5}=0.8$~pc (NMS8OP model) 
eject their $\rmn{S_{MAXI}}$, while 8 (2) out of 100 (10) mass-segregated clusters with 
$M_{\rmn{ecl}}=10^{3.5}~(10^4)~\rmn{M_{\sun}}$ and $r_{0.5}=0.3$~pc (MS3OP model) 
eject their $\rmn{S_{MAXI}}$ (Table~3). In reality, many young star clusters are 
observed to fulfil the conditions which are needed to eject their massive star, 
such as having a compact size and massive binaries with similar component masses, 
so that some (up to $\approx 75$ per cent for $10^5~\rmn{M}_{\sun}$ clusters, see Fig.~4) 
of the real embedded clusters could have lost their initially most massive star 
by dynamical ejection within 3~Myr.

In this paper, we are only interested in the heaviest star of the cluster.
However the dynamical ejection of other massive stars in the cluster is also interesting
because it is important to understand the origin of field massive stars and OB runaways 
\citep{FP11} to help us to constrain the initial configuration of massive stars in clusters. 
The dynamical ejections of OB stars in our theoretical young star cluster library will 
be discussed in the following paper (Oh et al. in prep.), and \citet{BKO12} and \citet{PK06} 
have, respectively, already demonstrated that R136-type star-burst and ONC-type clusters 
are very efficient in ejecting OB stars.

\section*{Acknowledgments}
We are very grateful to Sverre Aarseth for making his {\sevensize NBODY}6 freely available 
and continuing its improvements, and to Vasilii Gvaramadze for useful comments. 
S. Oh was supported for this research through a stipend from the International
Max Planck Research School (IMPRS) for Astronomy and Astrophysics at the
Universities of Bonn and Cologne as well as by a stipend from the Stellar Populations 
and Dynamics Research (SPODYR) Group at the AIfA and from the DAAD.

\appendix
\section[]{Ejection estimates of $\rmn{S_{MAXI}}$ using the half-mass radius}
The tidal radius is used in the paper to determine if a star belongs to the cluster. 
In some cases, even though $\rmn{S_{MAXI}}$ is dynamically ejected through a strong close encounter 
the star may have not left the cluster yet at a given time, especially if it has obtained 
a relatively low velocity from the encounter. 
Thus the real probability for ejection of $\rmn{S_{MAXI}}$ may be higher than that presented in this paper.
Considering that the massive stars are generally seated in the central part of the cluster due to dynamical interactions,
relatively small distance such as the half-mass radius can be used as the ejection criterion for $\rmn{S_{MAXI}}$. 
However, it should be noted that $\rmn{S_{MAXI}}$ being found outside of the half-mass radius
does not necessarily mean that is ejected, especially when the cluster is initially unsegregated
and dynamically unevolved.
Here, therefore, we provide tables (Table~A1 and A2) which contain the mean half-mass radius and 
the numbers of clusters whose $\rmn{S_{MAXI}}$ is found outside of $r_{0.5}$, $2r_{0.5}$ and $4r_{0.5}$
and additionally has a velocity greater than the escape velocity at the given time.
For most of the cluster models the mean half-mass radii do not change much within 3~Myr due to 
the clusters' dynamical evolution. Note that the initial crossing time of the clusters with
$M_{\rmn{ecl}}\leq100~\rmn{M}_{\sun}$ and $r_{0.5}=0.8$~pc is longer than 3~Myr (Table~2).
The most expanded half-mass radius is $\approx$ 1~pc at 3~Myr, although a few massive cluster models have
expanded up to twice their initial half-mass radius.

The initial positions of $\rmn{S_{MAXI}}$ distinctly show whether the models are generated with initial mass segregation. 
For the initially unsegregated clusters (Table~A1), more than half of the realizations for each models show that 
the $\rmn{S_{MAXI}}$ is initially located outside of the half-mass radius. 
Even in up to 23 (8) per~cent of the clusters the star is located further out than $2r_{0.5}$ ($4r_{0.5}$) at 0~Myr.
For the initially mass-segregated clusters, only in few low-mass clusters (less than 5 per cent at most) is
$\rmn{S_{MAXI}}$ located outside of $r_{0.5}$ at 0~Myr (Table~A2). 
It may be strange for the mass-segregated cluster to show its most massive star 
having an initial position beyond the half-mass radius at all. But for the initially mass-segregated 
clusters with low-masses ($M_{\rmn{ecl}}<100~\rmn{M}_{\sun}$) the initial position of $\rmn{S_{MAXI}}$ can, 
sometimes, be generated slightly beyond the half-mass radius due to the shallow gravitational potential, 
the low number statistics and the algorithm generating the initially mass-segregated cluster 
(the heaviest star being the {\it most bound} to the cluster, but not requiring it to be at the most central position). 
None of the initially mass-segregated clusters have $\rmn{S_{MAXI}}$ located initially 
further out than $2 r_{0.5}$.

For the initially not mass-segregated clusters with relatively high masses ($\geq100~\rmn{M}_{\sun}$) 
and/or $r_{0.5}=0.3~\rmn{pc}$, $N(r_{\rmn{S_{MAXI}}}>r_{0.5})$ in Table~A1 significantly decrease at 3~Myr
as a result of dynamical evolution, i.e. dynamical mass segregation. 
But there are still a large number of the unsegregated clusters, particularly the low-density ones, 
which have $\rmn{S_{MAXI}}$ beyond the half-mass radius. In many of the cases the star has a velocity 
lower than the escape velocity.  
As those clusters are dynamically young and the velocity of the star is too small to escape from its cluster,
it is unlikely that these stars are dynamically ejected. Either the star was initially located beyond 
the half-mass radius and has not fallen into the cluster centre yet. 
Or, if $\rmn{S_{MAXI}}$ of the low-density clusters has an initial orbit as large as the half-mass radius
(note that the half-mass radius is less than a pc), it could be temporarily found beyond the half-mass radius since
the dynamical interactions which lead to the massive star being confined to the central part of the cluster 
are insufficient by 3~Myr for these low-density clusters. 
This can also explain that for the mass-segregated clusters with low masses, especially with $10~\rmn{M}_{\sun}$, 
$N(r_{\rmn{S_{MAXI}}}>r_{0.5})$ increases at 3~Myr and most of these clusters have $\rmn{S_{MAXI}}$ moving slower 
than the escape velocity.
There are only two initially mass segregated clusters with $10~\rmn{M}_{\sun}$ whose $\rmn{S_{MAXI}}$ is 
found beyond twice the half-mass radius at 3~Myr.
There are due to low-energy encounters expelling the stars from the core without ejecting them from the cluster.
However, in the NMS3OP and MS3OP models, and in massive ($\geq 10^{3}~\rmn{M}_{\sun}$) clusters from the other models,  
a number of clusters have $\rmn{S_{MAXI}}$ being beyond the half-mass radius with a velocity 
larger than the escape velocity, which implies that the $\rmn{S_{MAXI}}$ stars are dynamically ejected. 
The probability of $\rmn{S_{MAXI}}$ ejection increases with cluster mass. 40 per cent of the $10^{4}~\rmn{M}_{\sun}$ 
MS3OP clusters have probably ejected their $\rmn{S_{MAXI}}$.

\begin{table*}
\centering
 \caption{The averaged half-mass radius, $<r_{0.5}>$, and the number of clusters whose $\rmn{S_{MAXI}}$ 
 is found beyond the half-mass radius, $N(r_{\rmn{S_{MAXI}}}>r_{0.5})$, beyond $2 r_{0.5}$, 
 $N(r_{\rmn{S_{MAXI}}}>2r_{0.5})$, and beyond $4 r_{0.5}$, $N(r_{\rmn{S_{MAXI}}}>4r_{0.5})$ at 0 and 3~Myr 
 from the initially unsegregated clusters. The latter numbers at 3~Myr marked with $^{\dag}$, are the number 
 of clusters whose $\rmn{S_{MAXI}}$ fulfils the distance criteria and moves faster than the escape velocity 
 ($v_{\rmn{S_{MAXI}}} > v_{\rmn{esc}}$). }
 \begin{tabular}{@{}lccccccccccccc@{}}
\hline
  $M_{\rmn{ecl}}[\rmn{M}_{\sun}]$ & \multicolumn{2}{c}{$<r_{0.5}>$~[pc]} & 
\multicolumn{3}{c}{$N(r_{\rmn{S_{MAXI}}}>r_{0.5})$} && \multicolumn{3}{c}{$N(r_{\rmn{S_{MAXI}}}>2r_{0.5})$} 
&& \multicolumn{3}{c}{$N(r_{\rmn{S_{MAXI}}}>4r_{0.5})$}\\
&0~Myr&3~Myr&0~Myr&3~Myr&3~Myr$^{\dag}$&&0~Myr&3~Myr&3~Myr$^{\dag}$&&0~Myr&3~Myr&3~Myr$^{\dag}$\\
\hline
\multicolumn{14}{@{}l}{NMS3S} \\
  $10$       & 0.27 & 0.27 & 49 & 31 & 0  && 14 & 6  & 0  && 2  & 1  & 0  \\
  $10^{1.5}$ & 0.29 & 0.30 & 47 & 6  & 0  && 9  & 3  & 0  && 2  & 0  & 0  \\
  $10^{2}$   & 0.30 & 0.36 & 52 & 11 & 0  && 21 & 8  & 0  && 3  & 4  & 0  \\
  $10^{2.5}$ & 0.30 & 0.41 & 54 & 6  & 0  && 16 & 3  & 0  && 3  & 0  & 0  \\
  $10^{3}$   & 0.30 & 0.50 & 45 & 2  & 0  && 15 & 2  & 0  && 7  & 1  & 0  \\
  $10^{3.5}$ & 0.30 & 0.55 & 48 & 10 & 6  && 18 & 8  & 6  && 3  & 5  & 4  \\
\multicolumn{14}{@{}l}{NMS3RP} \\
  $10$       & 0.25 & 0.31 & 53 & 23 & 1  && 6  & 0  & 0  && 0  & 0  & 0  \\
  $10^{1.5}$ & 0.28 & 0.31 & 46 & 12 & 0  && 9  & 3  & 0  && 1  & 1  & 0  \\
  $10^{2}$   & 0.29 & 0.37 & 53 & 17 & 0  && 21 & 10 & 0  && 4  & 3  & 0  \\
  $10^{2.5}$ & 0.30 & 0.47 & 54 & 12 & 0  && 19 & 1  & 0  && 3  & 0  & 0  \\
  $10^{3}$   & 0.30 & 0.54 & 47 & 6  & 2  && 17 & 3  & 1  && 7  & 1  & 0  \\
  $10^{3.5}$ & 0.30 & 0.55 & 50 & 13 & 6  && 18 & 9  & 6  && 3  & 7  & 6  \\
\multicolumn{14}{@{}l}{NMS3OP} \\
  $10$       & 0.26 & 0.30 & 46 & 23 & 0  && 12 & 6  & 0  && 1  & 1  & 0  \\
  $10^{1.5}$ & 0.28 & 0.31 & 44 & 18 & 0  && 12 & 4  & 0  && 0  & 1  & 0  \\
  $10^{2}$   & 0.29 & 0.38 & 47 & 9  & 2  && 18 & 4  & 1  && 1  & 0  & 0  \\
  $10^{2.5}$ & 0.30 & 0.56 & 44 & 14 & 7  && 19 & 8  & 3  && 5  & 3  & 2  \\
  $10^{3}$   & 0.30 & 0.68 & 44 & 15 & 11 && 16 & 9  & 9  && 6  & 7  & 7  \\
  $10^{3.5}$ & 0.30 & 0.65 & 48 & 17 & 15 && 15 & 16 & 15 && 3  & 14 & 14 \\
\multicolumn{14}{@{}l}{NMS8S} \\
  $10$       & 0.70 & 0.67 & 45 & 39 & 0  && 11 & 10 & 0  && 4  & 4  & 0  \\
  $10^{1.5}$ & 0.79 & 0.77 & 58 & 50 & 0  && 17 & 19 & 0  && 3  & 3  & 0  \\
  $10^{2}$   & 0.79 & 0.81 & 49 & 35 & 0  && 10 & 9  & 0  && 1  & 1  & 0  \\
  $10^{2.5}$ & 0.80 & 0.82 & 49 & 25 & 0  && 22 & 16 & 0  && 8  & 8  & 0  \\
  $10^{3}$   & 0.80 & 0.84 & 43 & 18 & 0  && 16 & 5  & 0  && 4  & 3  & 0  \\
  $10^{3.5}$ & 0.81 & 0.86 & 38 & 10 & 0  && 16 & 3  & 0  && 3  & 1  & 0  \\
\multicolumn{14}{@{}l}{NMS8RP}\\
  $10$       & 0.68 & 0.67 & 51 & 46 & 0  && 12 & 12 & 0  && 0  & 1  & 0  \\
  $10^{1.5}$ & 0.74 & 0.75 & 46 & 46 & 0  && 18 & 16 & 0  && 3  & 3  & 0  \\
  $10^{2}$   & 0.79 & 0.82 & 43 & 34 & 0  && 15 & 13 & 0  && 1  & 1  & 0  \\
  $10^{2.5}$ & 0.80 & 0.82 & 51 & 20 & 0  && 16 & 13 & 0  && 2  & 1  & 0  \\
  $10^{3}$   & 0.80 & 0.84 & 41 & 13 & 0  && 14 & 9  & 0  && 4  & 4  & 0  \\
  $10^{3.5}$ & 0.80 & 0.85 & 52 & 18 & 0  && 15 & 7  & 0  && 5  & 2  & 0  \\
\multicolumn{14}{@{}l}{NMS8OP} \\
  $10$       & 0.69 & 0.68 & 39 & 42 & 0  && 8  & 11 & 0  && 0  & 0  & 0  \\
  $10^{1.5}$ & 0.76 & 0.76 & 48 & 34 & 0  && 10 & 10 & 0  && 1  & 2  & 0  \\
  $10^{2}$   & 0.77 & 0.81 & 45 & 20 & 0  && 10 & 7  & 0  && 2  & 2  & 0  \\
  $10^{2.5}$ & 0.79 & 0.85 & 63 & 32 & 0  && 23 & 13 & 0  && 6  & 6  & 0  \\
  $10^{3}$   & 0.81 & 0.88 & 53 & 18 & 2  && 18 & 8  & 2  && 2  & 4  & 1  \\
  $10^{3.5}$ & 0.80 & 0.93 & 44 & 12 & 1  && 17 & 3  & 0  && 0  & 2  & 0  \\
 \hline
 \end{tabular}
\end{table*}

\begin{table*}
\centering 
\caption{Same as Table~A1 but for the initially mass-segregated cluster models. Note that only 10 realizations are 
 performed for $10^{4}~\rmn{M}_{\sun}$ clusters due to the expensive computing cost. }
 \begin{tabular}{@{}lccccccccccccc@{}}
\hline
  $M_{\rmn{ecl}}[\rmn{M}_{\sun}]$ & \multicolumn{2}{c}{$<r_{0.5}>$~[pc]} &
\multicolumn{3}{c}{$N(r_{\rmn{S_{MAXI}}}>r_{0.5})$} && \multicolumn{3}{c}{$N(r_{\rmn{S_{MAXI}}}>2r_{0.5})$} 
&& \multicolumn{3}{c}{$N(r_{\rmn{S_{MAXI}}}>4r_{0.5})$}\\
&0~Myr&3~Myr&0~Myr&3~Myr&3~Myr$^{\dag}$&&0~Myr&3~Myr&3~Myr$^{\dag}$&&0~Myr&3~Myr&3~Myr$^{\dag}$\\
 \hline
 \multicolumn{14}{@{}l}{MS3S}\\
  $10$       & 0.24 & 0.25 & 0 & 10 & 0  && 0  & 0  & 0  && 0  & 0  & 0 \\
  $10^{1.5}$ & 0.28 & 0.31 & 0 & 0  & 0  && 0  & 0  & 0  && 0  & 0  & 0 \\
  $10^{2}$   & 0.32 & 0.35 & 0 & 0  & 0  && 0  & 0  & 0  && 0  & 0  & 0 \\
  $10^{2.5}$ & 0.31 & 0.44 & 0 & 2  & 1  && 0  & 0  & 0  && 0  & 0  & 0 \\
  $10^{3}$   & 0.31 & 0.55 & 0 & 2  & 1  && 0  & 2  & 1  && 0  & 1  & 1 \\
  $10^{3.5}$ & 0.31 & 0.59 & 0 & 7  & 5  && 0  & 5  & 4  && 0  & 4  & 4 \\
 \multicolumn{14}{@{}l}{MS3RP} \\
  $10$       & 0.23 & 0.29 & 4 & 25 & 1  && 0  & 2  & 0  && 0  & 0  & 0 \\
  $10^{1.5}$ & 0.29 & 0.30 & 3 & 2  & 0  && 0  & 0  & 0  && 0  & 0  & 0 \\
  $10^{2}$   & 0.30 & 0.39 & 0 & 3  & 0  && 0  & 0  & 0  && 0  & 0  & 0 \\
  $10^{2.5}$ & 0.31 & 0.47 & 0 & 3  & 3  && 0  & 1  & 1  && 0  & 1  & 1 \\
  $10^{3}$   & 0.31 & 0.57 & 0 & 3  & 3  && 0  & 3  & 3  && 0  & 2  & 2 \\
  $10^{3.5}$ & 0.31 & 0.59 & 0 & 11 & 6  && 0  & 9  & 6  && 0  & 8  & 6 \\
 \multicolumn{14}{@{}l}{MS3OP} \\
  $10$       & 0.23 & 0.26 & 2 & 11 & 0  && 0  & 0  & 0  && 0  & 0  & 0 \\
  $10^{1.5}$ & 0.28 & 0.33 & 0 & 3  & 1  && 0  & 1  & 1  && 0  & 0  & 0 \\
  $10^{2}$   & 0.29 & 0.42 & 0 & 4  & 4  && 0  & 4  & 4  && 0  & 2  & 2 \\
  $10^{2.5}$ & 0.30 & 0.60 & 0 & 19 & 16 && 0  & 12 & 12 && 0  & 3  & 3 \\
  $10^{3}$   & 0.31 & 0.72 & 0 & 20 & 13 && 0  & 15 & 11 && 0  & 11 & 8 \\
  $10^{3.5}$ & 0.31 & 0.70 & 0 & 26 & 22 && 0  & 24 & 22 && 0  & 19 & 19\\
  $10^{4}$   & 0.31 & 0.50 & 0 & 4  & 4  && 0  & 4  & 4  && 0  & 4  & 4 \\
 \multicolumn{14}{@{}l}{MS8S} \\
  $10$       & 0.63 & 0.59 & 0 & 6 & 0  && 0  & 0  & 0  && 0  & 0  & 0 \\
  $10^{1.5}$ & 0.77 & 0.74 & 1 & 5  & 0  && 0  & 0  & 0  && 0  & 0  & 0 \\
  $10^{2}$   & 0.87 & 0.82 & 0 & 1  & 0  && 0  & 0  & 0  && 0  & 0  & 0 \\
  $10^{2.5}$ & 0.81 & 0.86 & 0 & 0  & 0  && 0  & 0  & 0  && 0  & 0  & 0 \\
  $10^{3}$   & 0.82 & 0.90 & 0 & 0  & 0  && 0  & 0  & 0  && 0  & 0  & 0 \\
  $10^{3.5}$ & 0.83 & 0.93 & 0 & 0  & 0  && 0  & 0  & 0  && 0  & 0  & 0 \\
 \multicolumn{14}{@{}l}{MS8RP} \\
  $10$       & 0.60 & 0.57 & 3 & 3  & 0  && 0  & 0  & 0  && 0  & 0  & 0 \\
  $10^{1.5}$ & 0.74 & 0.70 & 0 & 8 & 0  && 0  & 0  & 0  && 0  & 0  & 0 \\
  $10^{2}$   & 0.81 & 0.82 & 0 & 3  & 0  && 0  & 0  & 0  && 0  & 0  & 0 \\
  $10^{2.5}$ & 0.82 & 0.85 & 0 & 2  & 0  && 0  & 0  & 0  && 0  & 0  & 0 \\
  $10^{3}$   & 0.83 & 0.90 & 0 & 0  & 0  && 0  & 0  & 0  && 0  & 0  & 0 \\
  $10^{3.5}$ & 0.83 & 0.95 & 0 & 0  & 0  && 0  & 0  & 0  && 0  & 0  & 0 \\
 \multicolumn{14}{@{}l}{MS8OP} \\
  $10$       & 0.61 & 0.55 & 0 & 3 & 0  && 0  & 0  & 0  && 0  & 0  & 0 \\
  $10^{1.5}$ & 0.75 & 0.67 & 0 & 4  & 0  && 0  & 0  & 0  && 0  & 0  & 0 \\
  $10^{2}$   & 0.78 & 0.80 & 0 & 2  & 0  && 0  & 0  & 0  && 0  & 0  & 0 \\
  $10^{2.5}$ & 0.80 & 0.93 & 0 & 3  & 2  && 0  & 0  & 0  && 0  & 0  & 0 \\
  $10^{3}$   & 0.83 & 1.02 & 0 & 7  & 4  && 0  & 2  & 2  && 0  & 0  & 0 \\
  $10^{3.5}$ & 0.83 & 1.05 & 0 & 5  & 3  && 0  & 3  & 2  && 0  & 2  & 2 \\
 \hline
 \end{tabular}
\end{table*}

\label{lastpage}

\end{document}